\documentclass{aa}  
\usepackage{graphicx}
\usepackage{subcaption}
\usepackage{amssymb}
\usepackage{amsmath}
\usepackage{txfonts}
\usepackage{gensymb}
\usepackage{hyperref}
\usepackage[normalem]{ulem}
\bibpunct{(}{)}{;}{a}{}{,}

\usepackage{color}
\definecolor{darkblue}{rgb}{0,0,0.6}
\newcommand{\kibitz}[2]{\ifnum\Comments=1\textcolor{#1}{#2}\fi}

\authorrunning{Magee \& Maguire}
\titlerunning{An investigation of $^{56}$Ni shells}

\begin{document} 
            
    \title{An investigation of $^{56}$Ni shells as the source of \\early light curve bumps in type Ia supernovae
} 
	\author{M. R. Magee \inst{1,2}
	\and 
	K. Maguire \inst{1}
	}

	\institute{School of Physics, Trinity College Dublin, The University of Dublin, Dublin 2, Ireland \\  (\email{mrmagee.astro@gmail.com} \label{inst1})
	\and Astrophysics Research Centre, School of Mathematics and Physics, Queen's University Belfast, Belfast, BT7 1NN, UK \label{inst2}
		}

   \date{Received -	- -; accepted - - - }

 
  \abstract{An excess of flux (i.e. a bump) in the early light curves of type Ia supernovae has been observed in a handful of cases. Multiple scenarios have been proposed to explain this excess flux. Recently, it has been shown that for at least one object (SN~2018oh) the excess emission observed could be the result of a large amount of $^{56}$Ni in the outer ejecta ($\sim$0.03~$M_{\rm{\odot}}$). We present a series of model light curves and spectra for ejecta profiles containing $^{56}$Ni shells of varying masses (0.01, 0.02, 0.03, and 0.04~$M_{\rm{\odot}}$) and widths. We find that even for our lowest mass $^{56}$Ni shell, an increase of \textgreater2 magnitudes is produced in the bolometric light curve at one day after explosion relative to models without a $^{56}$Ni shell. We show that the colour evolution of models with a $^{56}$Ni shell differs significantly from those without and shows a colour inversion similar to some double-detonation explosion models. Furthermore, spectra of our $^{56}$Ni shell models show that strong suppression of flux between $\sim$3\,700 -- 4\,000~$\AA$ close to maximum light appears to be a generic feature for this class of model. Comparing our models to observations of SNe~2017cbv and 2018oh, we show that a $^{56}$Ni shell of 0.02 -- 0.04~$M_{\rm{\odot}}$ can match shapes of the early optical light curve bumps, but the colour and spectral evolution are in disagreement. Our models also predict a strong UV bump that is not observed. This would indicate that an alternative origin for the flux excess is necessary. In addition, based on existing explosion scenarios, producing such a $^{56}$Ni shell in the outer ejecta as required to match the light curve shape, without the presence of additional short-lived radioactive material, may prove challenging. Given that only a small amount of $^{56}$Ni in the outer ejecta is required to produce a bump in the light curve, such non-monotonically decreasing $^{56}$Ni distributions in the outer ejecta must be rare, if they were to occur at all.
  }
\keywords{
	supernovae: general --- radiative transfer 
    }
   \maketitle
%
   
\section{Introduction}
\label{sect:intro}

Although there is general consensus that type Ia supernovae (SNe Ia) result from the thermonuclear explosions of white dwarfs in binary systems, there is currently little agreement on the types of systems required or the manner through which these systems explode as supernovae (see e.g. \citealt{livio--18, wang--18, jha--19, soker--19} for recent reviews of SNe~Ia). Many of the proposed scenarios broadly reproduce observations surrounding maximum light and therefore it is only through observations at either very late or very early times that differences between the various explosion and progenitor scenarios can be discerned. Particular attention has been paid to obtaining observations within hours to days of the explosion, as these times may show a `bump' in the light curve indicating interaction with a companion star \citep{kasen--10} or circumstellar material \citep{piro-16} or the presence of short-lived radioactive isotopes \citep{noebauer-17}.

\par

To date, only a handful of objects displaying such bumps have been observed, despite extensive observational effort. Although only a small number of examples are known, these objects show considerable diversity, and subsequently multiple scenarios have been invoked to explain the flux excesses. Although peculiar in its own right, iPTF14atg was the first SN Ia discovered that showed a flux excess at early times: a strong ultraviolet pulse was observed within the first $\sim$4 days following explosion. Based on the models of \cite{kasen--10}, \cite{cao--15} argue that this early UV emission was the result of interaction between the SN ejecta and companion star, and therefore iPTF14atg was produced via the single degenerate channel. In contrast, \cite{kromer--16} argue that the spectral evolution of iPTF14atg is incompatible with this scenario and is instead more similar to the violent merger of two white dwarfs; however, they note that this scenario in itself does not explain the early UV emission. Since the discovery of iPTF14atg, additional objects have been claimed to show early excess emission, including SN~2012cg \citep{marion--16}, SN~2016jhr \citep{jiang--2017}, and SN~2017cbv \citep{hosseinzadeh--17}. 

\par

\cite{hosseinzadeh--17} also compare their observations of SN~2017cbv to the interaction models of \cite{kasen--10} and find the models over-predict the UV luminosity at early times. Given this disagreement, they discuss the possibility that the early excess observed in SN~2017cbv is produced by a bubble of radioactive $^{56}$Ni. The resulting $^{56}$Ni distribution should have a noticeable impact on the early light curve and could produce the short-lived excess luminosity observed \citep{piro-nakar-2013, piro-nakar-2014, jiang--18, maeda--18}.  

\par

Indeed, the classical double-detonation explosion scenario (in which a helium shell is accumulated and ignites on the surface of a white dwarf) predicts large amounts of $^{56}$Ni and other short-lived radioactive isotopes in the outer ejecta \citep{livne--90, livne--91, woosley--94}. Previous studies have shown how the decay of these isotopes can produce light curve bumps similar to those observed \citep{noebauer-17, jiang--18}. \cite{maeda--18} investigate how the post-explosion composition of the helium shell (including the amount of $^{56}$Ni present) affects the model observables and can result in light curve bumps of varying strengths.

\par

Recently, the exceptionally high cadence light curve of SN~2018oh provided another example of an early flux excess \citep{li--19}. This bump is investigated in detail by both \cite{shappee--2019} and \cite{dimitriadis--19}, who consider multiple origins including companion or CSM interaction, the double detonation scenario, and the possibility of an irregular $^{56}$Ni distribution. \cite{shappee--2019} show that SN~2018oh is reasonably well matched up to $\sim$3 days after explosion by a model with a well mixed $^{56}$Ni distribution, while at later times less mixing is preferred. This would indicate that the $^{56}$Ni distribution does not decrease monotonically towards the outer ejecta. Furthermore, \cite{dimitriadis--19} show light curves from models for which there are two distinct $^{56}$Ni regions within the ejecta; the majority of $^{56}$Ni resides in the centre and powers the main light curve, while a separate shell (0.03~$M_{\rm{\odot}}$) of $^{56}$Ni in the outer ejecta powers the initial flux excess. \cite{dimitriadis--19} find favourable agreement with the light curve shape of this scenario, but prefer an interpretation of interaction based on the colour evolution. 

\par

Further motivation for the study of irregular $^{56}$Ni distributions comes from supernova remnants. Some SNe~Ia remnants show evidence for possible large clumps of iron in the outer ejecta (at these late epochs, radioactive $^{56}$Ni has decayed fully to stable $^{56}$Fe, e.g.~\citealt{tsebrenko--15, yamaguchi--17, sato--20}). In addition, \cite{sato--19} argue that the structure observed in X-ray images of SNe~Ia remnants results from an initially clumpy structure to the explosion itself. The effect of similar clumps on the early light curve however has not been fully explored.

\par

For the first time, we present spectra and light curves resulting from Chandrasekhar-mass models with a shell of $^{56}$Ni in the outer ejecta. We compare our radiative transfer models to SNe~2018oh and 2017cbv, for which non-monotonic $^{56}$Ni distributions have been considered as possible causes of the observed early excesses, and explore the parameter space that could provide reasonable agreement to both objects. In Sect.~\ref{sect:models} we outline the construction of our models. We compare the effects of our non-monotonic $^{56}$Ni to models with a smoothly varying $^{56}$Ni distribution in Sect.~\ref{sect:clump_effects} and to models with extended $^{56}$Ni distributions in Sect.~\ref{sect:clump_v_extended}. A comparison with observations is presented in Sect.~\ref{sect:comparisons} along with discussion in Sect.~\ref{sect:discussion}. Finally, we present our conclusions in Sect.~\ref{sect:conclusions}.

%

\section{Models}
\label{sect:models}

\subsection{Radiative-transfer modelling}
All model light curves and spectra presented in this work were calculated using TURTLS \citep{magee--18} and are available on GitHub\footnote{\href{https://github.com/MarkMageeAstro/TURTLS-Light-curves}{https://github.com/MarkMageeAstro/TURTLS-Light-curves}}. In what follows we provide a brief overview of TURTLS. We refer the reader to \cite{magee--18} for a full description of the code and \cite{magee--20} for an application. TURTLS is a one-dimensional Monte Carlo radiative transfer code designed for modelling the early time evolution of SNe Ia. The density and composition of the SN ejecta are taken as input parameters and freely defined by the user in a series of discrete cells. Monte Carlo packets representing bundles of photons are injected into the model, tracing the decay of $^{56}$Ni. The propagation of these packets is followed throughout the model ejecta for a series of logarithmically separated time steps. For the models presented in this work, simulations are calculated between 0.5 and 30 days after explosion. We limit our simulations to 30 days after explosion due to the assumption of local-thermodynamic equilibrium. Packets are initially injected as $\gamma$-packets, representing $\gamma$-ray photons, and are followed assuming a $\gamma$-ray opacity of 0.03~cm$^{2}$~g$^{-1}$. After an interaction with the ejecta, $\gamma$-packets are converted to $r$-packets, representing optical photons. To follow the propagation of $r$-packets, we use TARDIS \citep{tardis, tardis_v2} to calculate the non-grey expansion opacities in each cell during each time step and also include the effects of electron scattering. Packets emerging from the model region are binned in terms of the time and frequency with which they escaped to construct synthetic observables. Light curves are constructed via a convolution of emerging packet luminosities and the desired set of filter functions.

\begin{figure}
\centering
\includegraphics[width=\columnwidth]{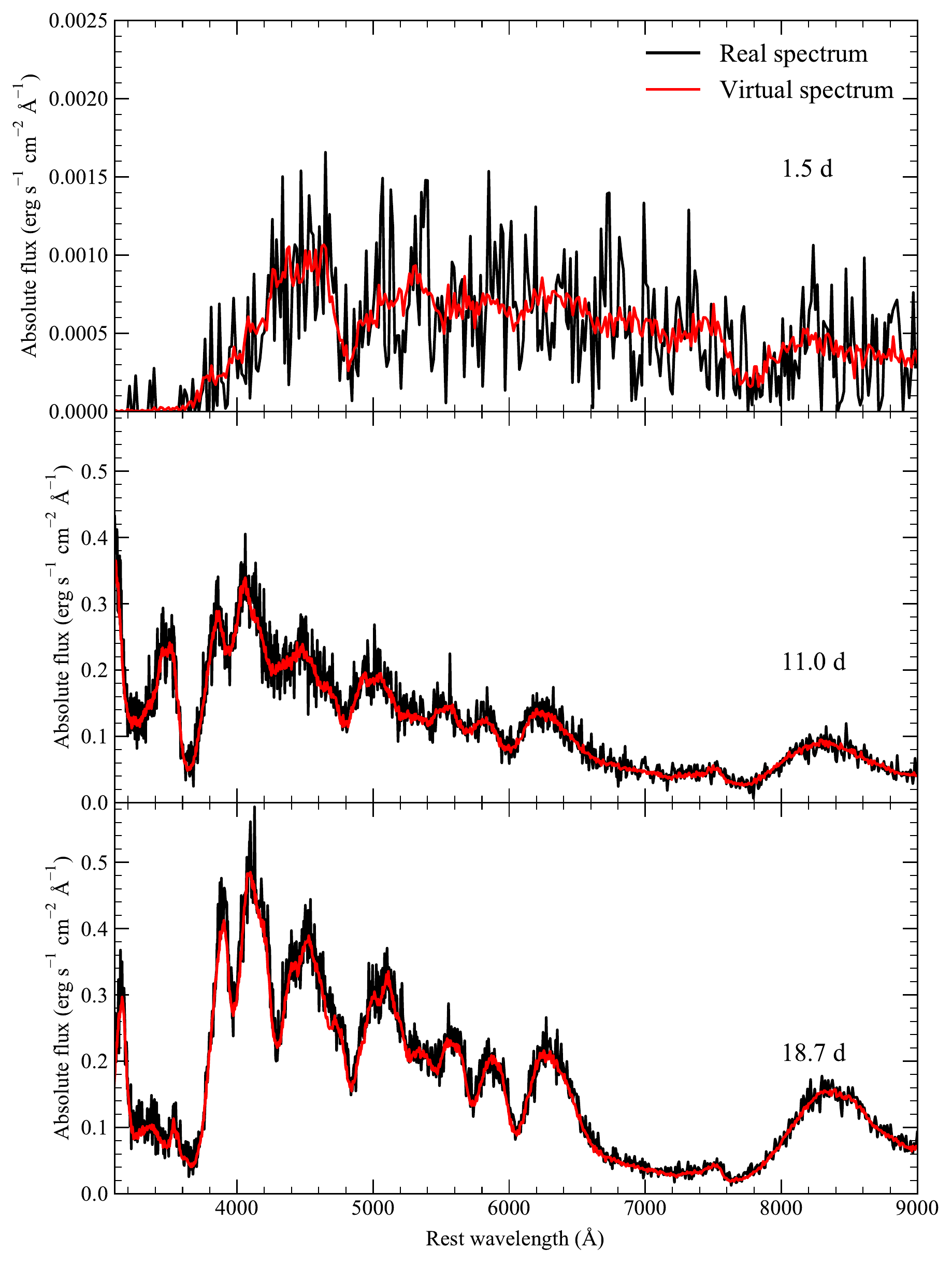}
\caption{Comparison of spectra calculated during different time steps (1.5\,d, 11.0\,d, and 18.7\,d) for our fiducial SN~2018oh model (see Sect.~\ref{sect:models}). Virtual spectra (red) are calculated using the event-based technique, while real spectra (black) are calculated by directly counting the luminosities of escaping real packets.}
\label{fig:virt}
\centering
\end{figure}

\par 

\begin{figure*}
\centering
\includegraphics[width=\textwidth]{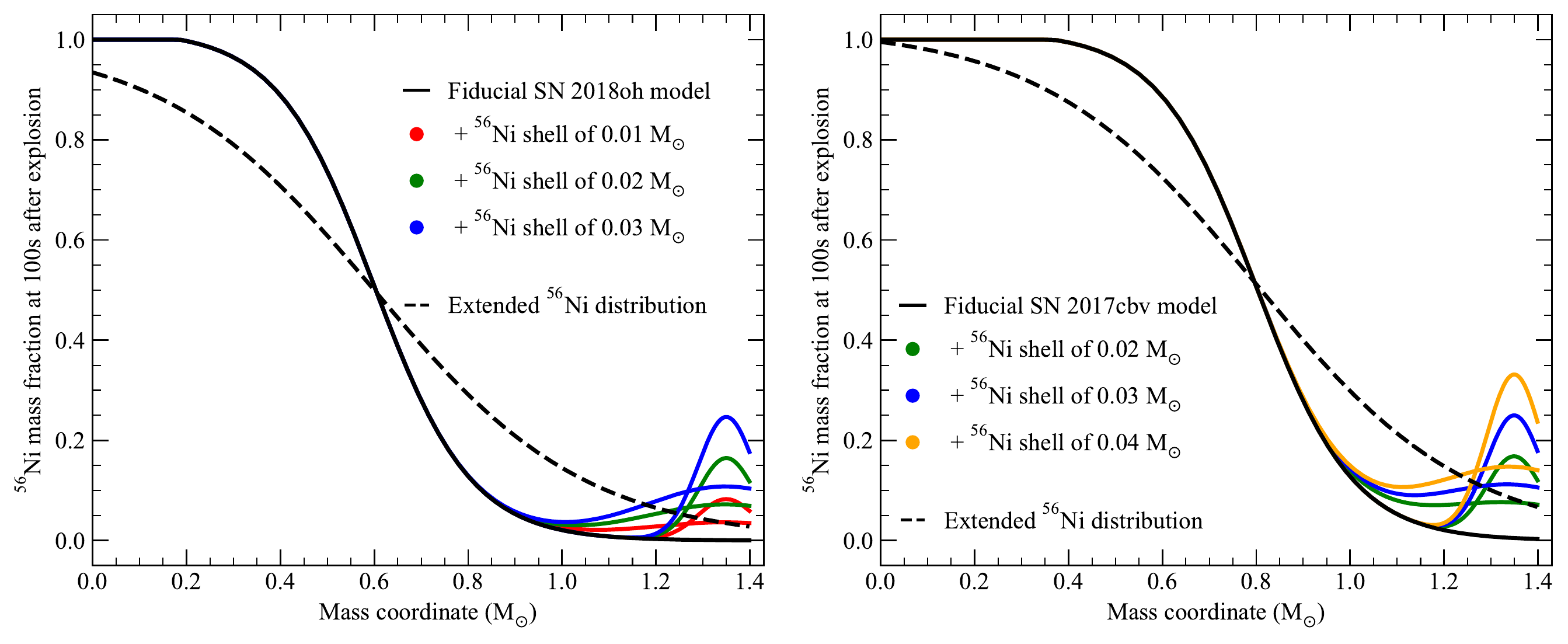}
\caption{$^{56}$Ni distributions for models explored as part of this work. Panels show the best matching models to the light curves beginning approximately five days after explosion (i.e. excluding the early bump) of SN~2018oh (left) and SN~2017cbv (right) as black solid lines, which we consider as our fiducial models. Dashed black lines show models that have more extended $^{56}$Ni distributions compared to our fiducial models, but are otherwise identical (i.e. the same density profile and $^{56}$Ni mass in each case). Coloured lines show $^{56}$Ni shells of varying masses. Each shell is centred on a mass coordinate of 1.35~$M_{\rm{\odot}}$ and has a width of either 0.06 or 0.18~$M_{\rm{\odot}}$. }
\label{fig:clump_distributions}
\centering
\end{figure*}

In order to improve the signal-to-noise ratio of our synthetic observables during the crucial early phases, we implemented into TURTLS the event-based technique (EBT) for spectrum extraction with virtual packets (see \citealt{long--knigge--02, sim--10-agn, tardis, bulla--15}). Briefly, in the EBT, after a Monte Carlo packet performs an interaction with the model ejecta, a given number of `virtual packets' ($v$-packets) are spawned. Each $v$-packet is emitted in a random direction with the same frequency in the fluid-frame as the real packets. Once injected, $v$-packets follow the same computational procedure as real packets, with the exception being that they do not interact with the ejecta. At time step $t$, an escaping $v$-packet contributes luminosity, $L_{\nu}$, to the spectrum, which is given by:
\begin{equation}
    L_{\nu} = \frac{E}{N\Delta t\Delta\nu} e^{-\tau},
\end{equation}
where $E$ is the energy of the real packet that spawned the $v$-packet, $N$ is the number of $v$-packets created for each real packet, $\Delta t$ is the width of the time step, $\Delta \nu$ is the width of the frequency bin, and $\tau$ is the total optical depth accumulated along the trajectory of the virtual packet. In this way, a `virtual' spectrum is created from these packets for each time step in the simulation. The virtual light curve at this time is then calculated by integrating the convolution of this spectrum and the filter functions. This method has previously been shown to reproduce the synthetic observables calculated by direct binning of the real packets (see e.g. Fig.~4 of \citealt{tardis} or Fig.~8 of \citealt{bulla--15}). In Fig.~\ref{fig:virt} we show the dramatic increase in the signal-to-noise ratio of our synthetic spectra using $v$-packets compared to the direct counting of emerging $r$-packets.

\par

\subsection{Constructing models with $^{56}$Ni shells}

\cite{dimitriadis--19} argue that the flux excess observed in the early light curve of SN~2018oh is consistent with a model containing two distinct regions of $^{56}$Ni in the ejecta. In this model, the main rise of the light curve is provided by the majority of $^{56}$Ni, which is located within the centre of the ejecta. An additional 0.03~$M_{\rm{\odot}}$ of $^{56}$Ni is placed near the surface of the ejecta (above a mass coordinate of 1.3~$M_{\rm{\odot}}$) and powers the initial excess. Following from \cite{dimitriadis--19}, we investigated the range of parameters that could plausibly reproduce the excesses observed in SNe~2018oh and 2017cbv.

\par

As a starting point, we used the models presented in \cite{magee--20}. One of the main limitations of these models is the composition, which uses a simplified three-zone structure (see Fig.~4 of \citealt{magee--18}). As the models were designed to explore solely the effects of the $^{56}$Ni distribution on the light curve, $^{56}$Ni constitutes 100\% of the iron-group element (IGE) zone (the centre of the ejecta) immediately after explosion. A small amount ($\sim$0.1~$M_{\rm{\odot}}$) of carbon and oxygen was placed in the outer ejecta, while the remaining mass was filled-in with intermediate-mass elements (IME). Due to these limitations, spectra in particular do not provide perfect agreement with observations, as they are more sensitive to specifics of the ejecta composition. The relatively large fraction of IMEs at high velocities for the models presented here is typically reflected in the spectra, with the corresponding features being shifted to higher velocities. Nevertheless, the differences between all models still represent a useful point of comparison. More physical and detailed composition treatments will be explored in future work. Furthermore, all models presented as part of this work have a total ejecta mass of 1.4~$M_{\rm{\odot}}$.

\par

To investigate the effects of $^{56}$Ni shells on the observables of both objects, we first found models from the existing set presented by \cite{magee--20} that produced reasonable agreement with the later light curves. We considered these the fiducial models and subsequently added $^{56}$Ni shells into the ejecta to determine whether this can reproduce the light curve bump. In determining the fiducial model, we followed the procedure outlined by \cite{magee--20}, but exclude  observations $\textgreater$14\,d before $B$-band maximum. Excluding these early points, we ignored the bump for both SNe and were only concerned with the main rise of the light curve, which will be primarily driven by the $^{56}$Ni produced in the core. For SN~2018oh, we find that the \cite{magee--20} model with 0.6~$M_{\rm{\odot}}$ of $^{56}$Ni, an exponential density profile, kinetic energy of 1.68$\times$10$^{51}$~erg, and intermediate $^{56}$Ni distribution ($s = 9.7$ in Eqn.~5 of \citealt{magee--20}) produced good agreement with the light curve beginning approximately five days after explosion. For SN~2017cbv, we find good agreement with a slightly higher $^{56}$Ni mass (0.8~$M_{\rm{\odot}}$) and lower kinetic energy (1.10$\times$10$^{51}$~erg). Taking these as our fiducial models, we added shells of pure $^{56}$Ni into the outer ejecta at different locations, varying the total $^{56}$Ni mass of the shell and its width.

\par

The general form of the $^{56}$Ni mass fraction within each shell follows a Gaussian distribution, with the $^{56}$Ni fraction at mass coordinate $m$ given by
\begin{equation}    
X_{\rm{Ni}}(m) = A \times e^{  -(m - \mu)^2 /{2\sigma^2}},
\end{equation}
where $\mu$ is the centre of the $^{56}$Ni shell in $M_{\rm{\odot}}$, $\sigma$ is the width of the shell in $M_{\rm{\odot}}$, and $A$ is a dimensionless scaling parameter used to control the total mass of $^{56}$Ni in the shell. This mass fraction is then added to the underlying fiducial model to give a description for the entire ejecta. The choice of a Gaussian distribution for the $^{56}$Ni mass fraction is arbitrary; however, we have also tested additional functional forms, such as those that are more similar to the inner regions of the ejecta (see Eqn.~11 of  \citealt{magee--18}). We find that the Gaussian distributions produce light curves that more closely resemble SNe~2018oh and SN~2017cbv, and therefore focus on these cases.  

\par

In Fig.~\ref{fig:clump_distributions} we show some of the model $^{56}$Ni distributions explored as part of this work. As shown in Fig.~\ref{fig:clump_distributions}, given that our models are in one dimension, the $^{56}$Ni shells would be similar to models containing large amounts of $^{56}$Ni in the outer ejecta only along a certain viewing angle. As each of our models has a corresponding fiducial model (i.e. a model without a $^{56}$Ni shell), we may consider the difference between the two to qualitatively reproduce the variation along different lines of sight. In this sense, our set of models broadly approximate the effect of a three-dimensional simulation with a large-scale clump of $^{56}$Ni visible along a specific line of sight (see Sect.~\ref{sect:discussion} for further discussion).

\par

We calculated light curves for $^{56}$Ni shells of 0.01, 0.02, 0.03, and 0.04~$M_{\rm{\odot}}$, with widths ($\sigma$) of 0.06 and 0.18~$M_{\rm{\odot}}$ and central locations of 1.35, 1.37, and 1.39~$M_{\rm{\odot}}$. In general, we find that changing the central location of the $^{56}$Ni shell had little effect on the overall light curve shape. Shells located further out in the ejecta (e.g. at 1.37 or 1.39~$M_{\rm{\odot}}$) produced earlier light curve bumps and redder colours beginning a few days after explosion, relative to shells that are somewhat deeper inside the ejecta (e.g. at 1.35~$M_{\rm{\odot}}$). The overall effect however, is secondary to the mass and width of the shell. We therefore focus on models with $^{56}$Ni shells centred on 1.35~$M_{\rm{\odot}}$ as these models produced the best agreement with observations (see Sect.~\ref{sect:comparisons}) and are representative of the general trends. 

\par

As a further point of comparison, we also include models with extended $^{56}$Ni distributions relative to the fiducial models. A key feature of our more extended $^{56}$Ni distribution models is that the $^{56}$Ni mass fraction is always monotonically decreasing towards the outer ejecta. This is clearly not the case for our $^{56}$Ni shell models, which have decreasing $^{56}$Ni mass fractions below 1.35~$M_{\rm{\odot}}$ before increasing again around $\sim1~M_{\rm{\odot}}$.

%

\section{Model observables}
\label{sect:model_lightcurves}

\subsection{Effects of $^{56}$Ni shells}
\label{sect:clump_effects}

\begin{figure*}
\centering
\includegraphics[width=\textwidth]{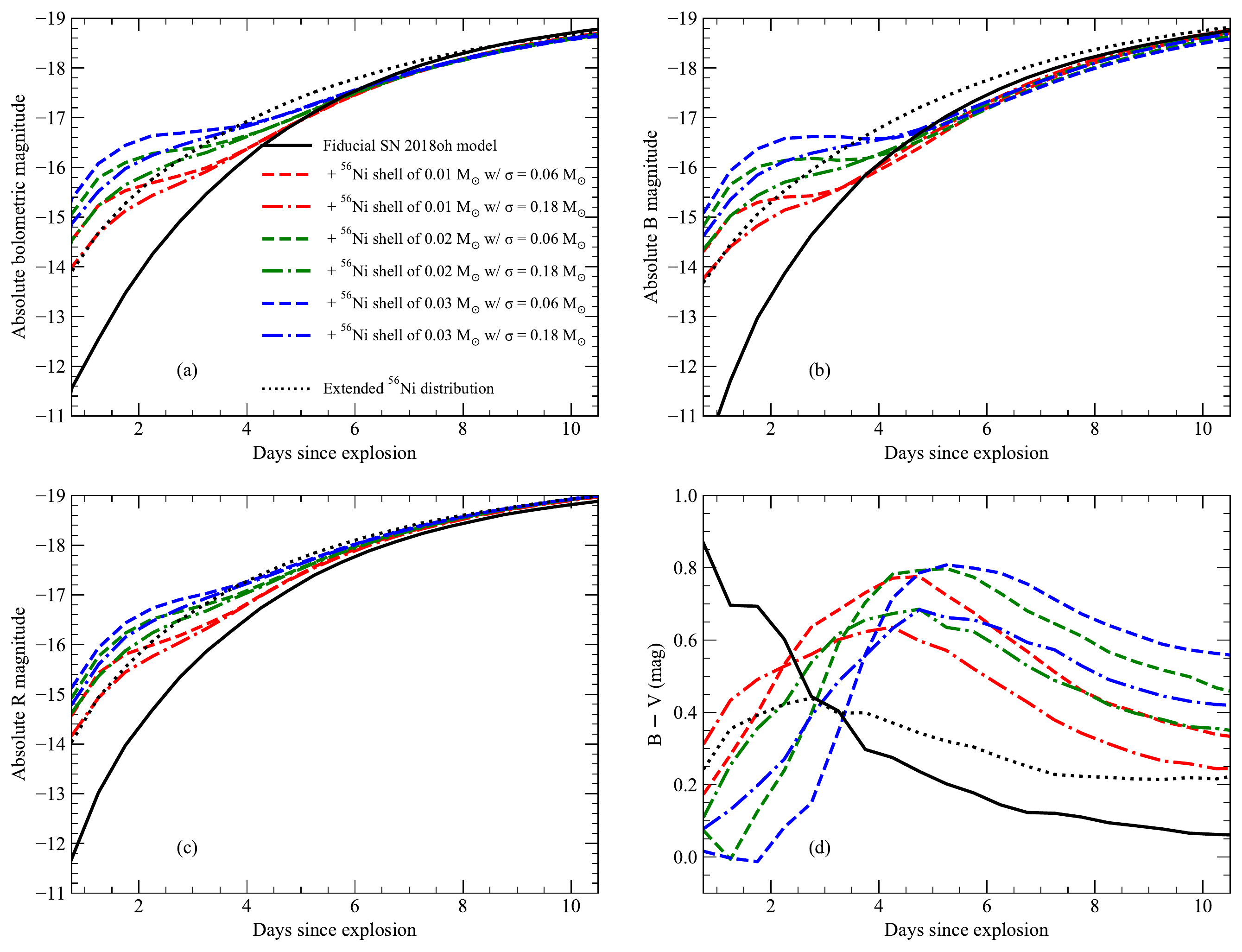}
\caption{Light curves for models with and without shells of $^{56}$Ni in the outer ejecta. Panels (a), (b), and (c) show the bolometric, $B$-, and $R$-band magnitudes, respectively. The $B-V$ colour evolution is shown in Panel (d). We determine our fiducial model for SN~2018oh by finding the best match to the later light curve (excluding the early bump) among the models of \cite{magee--20}. This is shown as a black solid line. Coloured lines show light curves resulting from models with added shells of $^{56}$Ni in the outer ejecta, of varying masses. Each shell is centred on a mass coordinate of 1.35~$M_{\rm{\odot}}$ and has a width of either 0.06 or 0.18~$M_{\rm{\odot}}$. Dashed black lines show a model that has a more extended $^{56}$Ni distributions compared to our fiducial model, but is otherwise identical (i.e. the same density profile and $^{56}$Ni mass in each case). Our models demonstrate that bumps in the early light curves, of varying sizes and lengths, can be generated by different shells of $^{56}$Ni in the outer ejecta.}
\label{fig:18oh_clump_lc}
\centering
\end{figure*}

In the following section we discuss generally the impact of $^{56}$Ni shells on the synthetic observables of the models. Here, we limit this discussion to a comparison between those models with $^{56}$Ni shells and those without. In Sect.~\ref{sect:clump_v_extended}, we discuss these models further alongside comparisons to a model with an extended $^{56}$Ni distribution relative to our fiducial SN~2018oh model. In Sect.~\ref{sect:comparisons}, we compare our shell models to observations of SNe~2018oh and 2017cbv and further discuss the implications for their specific explosion scenarios. We focus our discussion here on models in which shells have been added to the fiducial model for SN~2018oh (EXP\_Ni0.6\_KE1.68\_P9.7; see \citealt{magee--20} for further details on the model naming scheme). We find similar trends for models based on the SN~2017cbv fiducial model (EXP\_Ni0.8\_KE1.10\_P9.7). Both of our fiducial models represent intermediate values for the density profile and $^{56}$Ni distribution among the full parameter space explored in \cite{magee--20}.

\par

Figures~\ref{fig:18oh_clump_lc}(a), (b), and (c) demonstrate the effect of $^{56}$Ni shells on the bolometric and $BR$ band magnitudes compared to models without shells. We find that while the small increase in total $^{56}$Ni mass does not strongly affect the peak absolute magnitude of the models, there are significant differences during the first few days after explosion. For all of the parameters investigated here, an excess of flux in the light curve (relative to the fiducial model) is produced during this time -- the duration and total amount of flux excess varies. Again, we note that although our simulations are all performed in one dimension, differences between models with shells and our fiducial models are qualitatively similar to the variation in observables that may be expected along different lines of sight for a multi-dimensional model containing a clump only along a specific viewing angle.

\par

We find that in all cases of $^{56}$Ni shells, the flux excess is most easily observed in the bluer bands -- a natural consequence of the higher temperatures produced in the outer ejecta due to the presence of additional $^{56}$Ni heating. For example, in the case of our models with a 0.03~$M_{\rm{\odot}}$ $^{56}$Ni shell, the $B$-band is $\gtrsim$4 mag. brighter relative to our fiducial model at approximately one day after explosion. The $R$-band shows a slightly more modest increase of $\sim$3 magnitudes. Even for our lowest $^{56}$Ni mass shell (0.01~$M_{\rm{\odot}}$), the light curves are brighter relative to the fiducial model by $\sim$3 -- 3.5 mag. and $\sim$2 -- 2.5 mag. in the $B$- and $R$-bands, respectively. Aside from the total $^{56}$Ni mass contained within the shell, the distribution also affects the shape and colour of the flux excess. As shown in Fig.~\ref{fig:18oh_clump_lc}, those models with narrower $^{56}$Ni shells ($\sigma$ = 0.06~$M_{\rm{\odot}}$) typically show larger and bluer excesses over those models containing broader $^{56}$Ni shells ($\sigma$ = 0.18~$M_{\rm{\odot}}$). In addition, they show a somewhat more well-defined peak or bump -- broader $^{56}$Ni shells generally result in broader light curves with a `shoulder' rather than a distinct bump.

\par

For the models presented here, the duration of the flux excess ranges from $\sim$4 -- 5 days following explosion. Although the excess may have subsided at later times, the implications for the rest of the time evolution are profound and in particular the colour evolution for models with $^{56}$Ni shells differs significantly from those without, as shown in Fig.~\ref{fig:18oh_clump_lc}(d). For our fiducial model, the initial colour at one day after explosion is quite red ($B-V$ $\sim$ 0.8~mag.) due to a lack of $^{56}$Ni in the outer ejecta, but gradually becomes steadily bluer over the next week. Between 7 and 10 days after explosion, the $B-V$ colour has flattened and remains roughly constant at $B-V$ $\sim$ 0.1. In contrast, models with $^{56}$Ni shells show bluer colours (by $\sim$0.3 -- 0.5 mag.) within the first few days after explosion. By $\sim$4 -- 5 days following explosion, the $^{56}$Ni shell models have become significantly redder than the fiducial model, by as much as $\sim$0.6~mag. The $^{56}$Ni shell models subsequently become bluer again, but remain consistently redder than the fiducial model -- ranging from $\sim$0.1 -- 0.4 mag. redder around maximum light. 

\par

\begin{figure*}
\centering
\includegraphics[width=\textwidth]{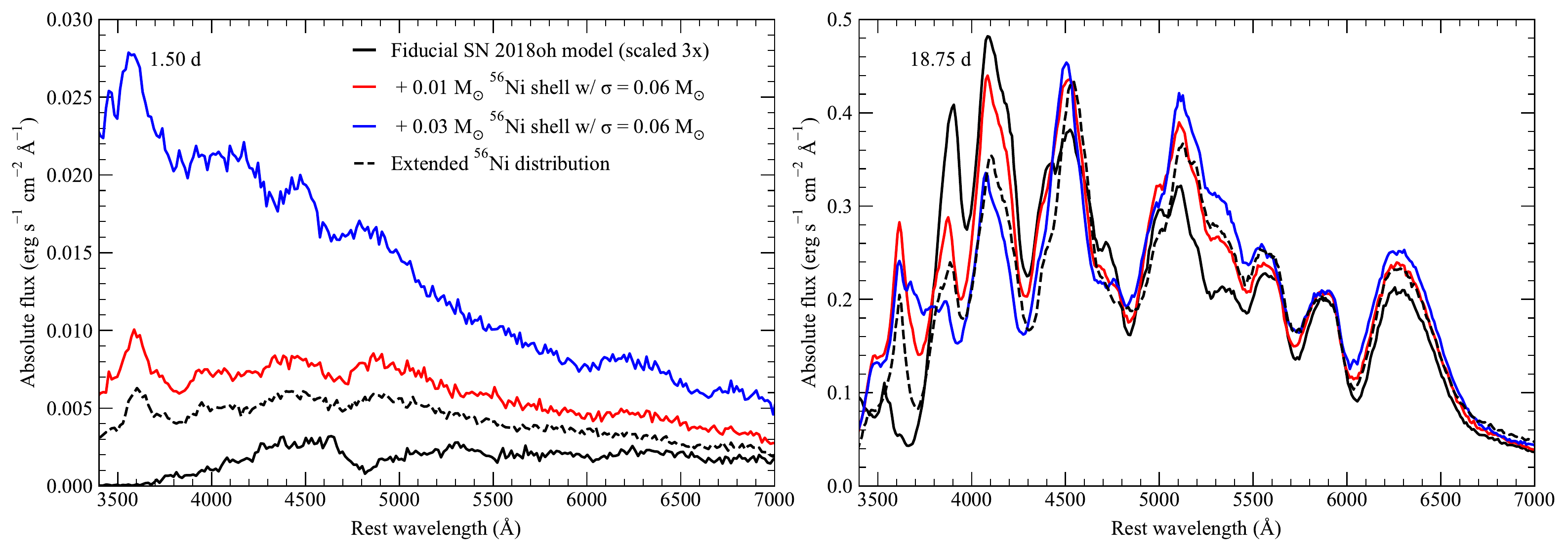}
\caption{Spectra for models with and without shells of $^{56}$Ni in the outer ejecta. {\it Left: } Spectra are shown for 1.50\,d after explosion, which corresponds to during the flux excess phase. All spectra have been binned to $\Delta\lambda$ = 15~\AA. We note that the spectrum of our fiducial model has been artificially scaled in flux by a factor of three. {\it Right: } Spectra are shown for 18.75\,d after explosion, corresponding to close to maximum light. All spectra have been binned to $\Delta\lambda$ = 10~\AA.}
\label{fig:clump_spectra}
\centering
\end{figure*}

In Fig.~\ref{fig:clump_spectra} we show spectra for our models at 1.50\,d and 18.75\,d after explosion. These phases correspond to approximately near the peak of the early flux excess and close to maximum light. We focus here on the case of our fiducial SN~2018oh model, models with $^{56}$Ni shells of 0.01 and 0.03~$M_{\rm{\odot}}$, shell widths of $\sigma$ = 0.06~$M_{\rm{\odot}}$, and a more extended $^{56}$Ni distribution relative to the fiducial model. As expected from the differences in light curves and colours at early times (Fig.~\ref{fig:18oh_clump_lc}), Fig.~\ref{fig:clump_spectra} shows that the spectra for our 0.03~$M_{\rm{\odot}}$ $^{56}$Ni shell model are significantly brighter and bluer than all other models at 1.50\,d after explosion. At this time, the light curve of our 0.01~$M_{\rm{\odot}}$ model is not unlike that of the extended $^{56}$Ni distribution model, hence their spectra are also quite similar. In contrast, the lack of $^{56}$Ni in the outer ejecta of our fiducial model produces a relatively flat and red spectrum, with the only prominent absorption feature appearing around $\sim$4\,800~\AA. We note that in Fig.~\ref{fig:clump_spectra}, the 1.50\,d spectrum of our fiducial model is scaled up by a factor of three to highlight the differences in spectroscopic features. 

\par

\begin{figure*}[h!]
    \centering
    \begin{subfigure}[b]{0.49\textwidth}
        \hspace*{-0.35cm}\includegraphics[height=5.85cm]{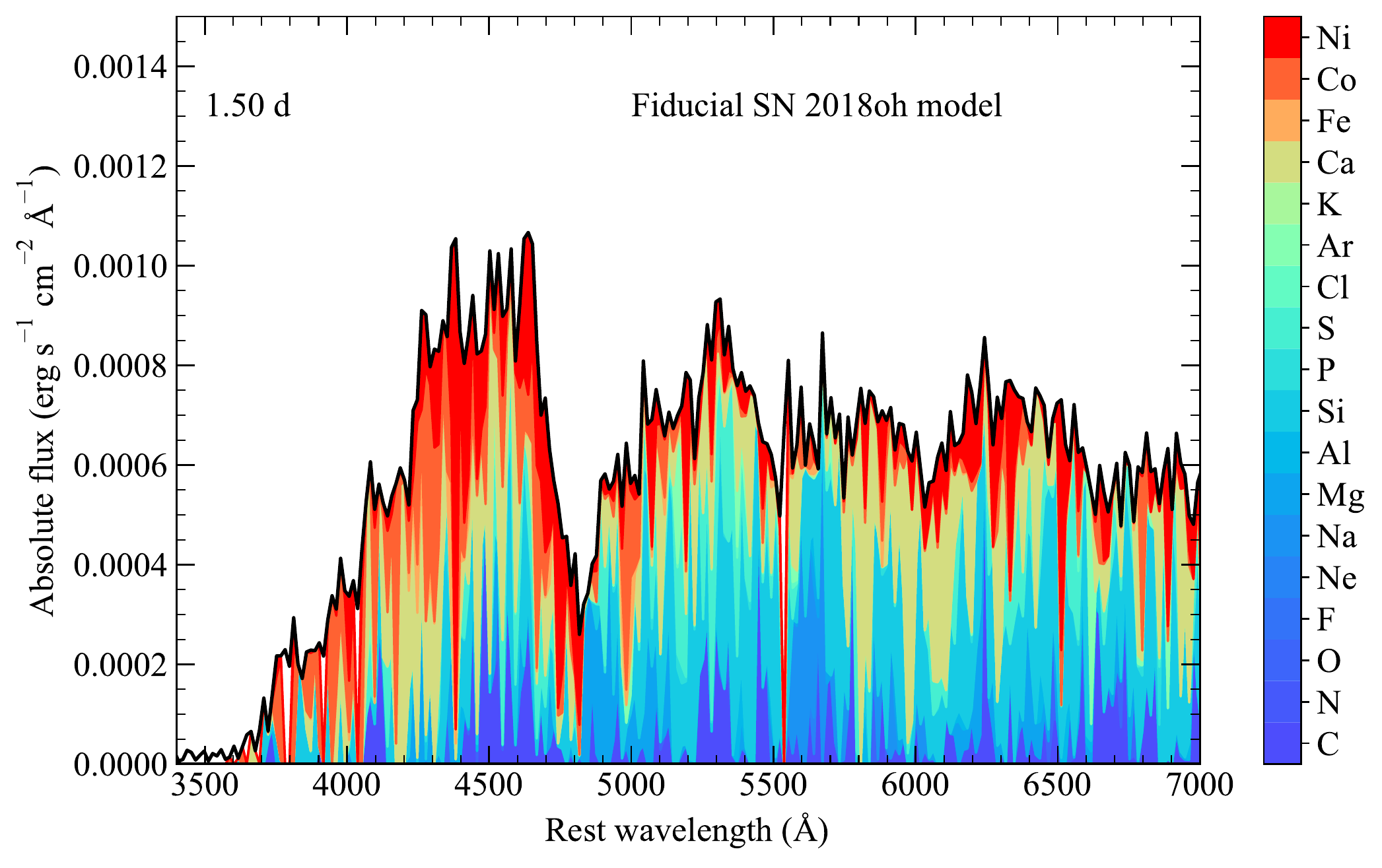}
    \end{subfigure}
    \begin{subfigure}[b]{0.49\textwidth}
        \hspace*{-0.18cm}\includegraphics[height=5.85cm]{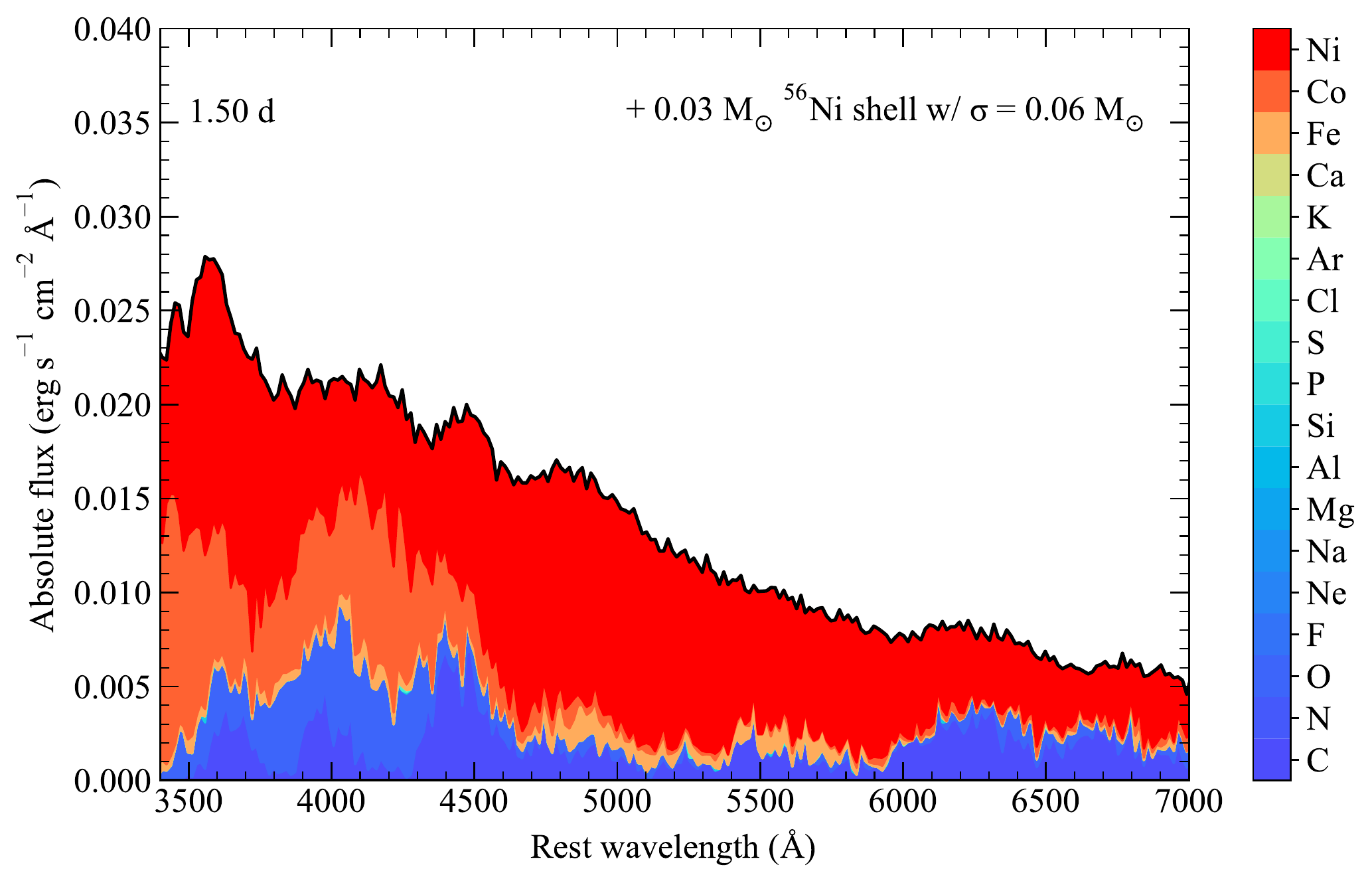}
    \end{subfigure} 
    \\
    \begin{subfigure}[b]{0.49\textwidth}
        \includegraphics[height=5.9cm]{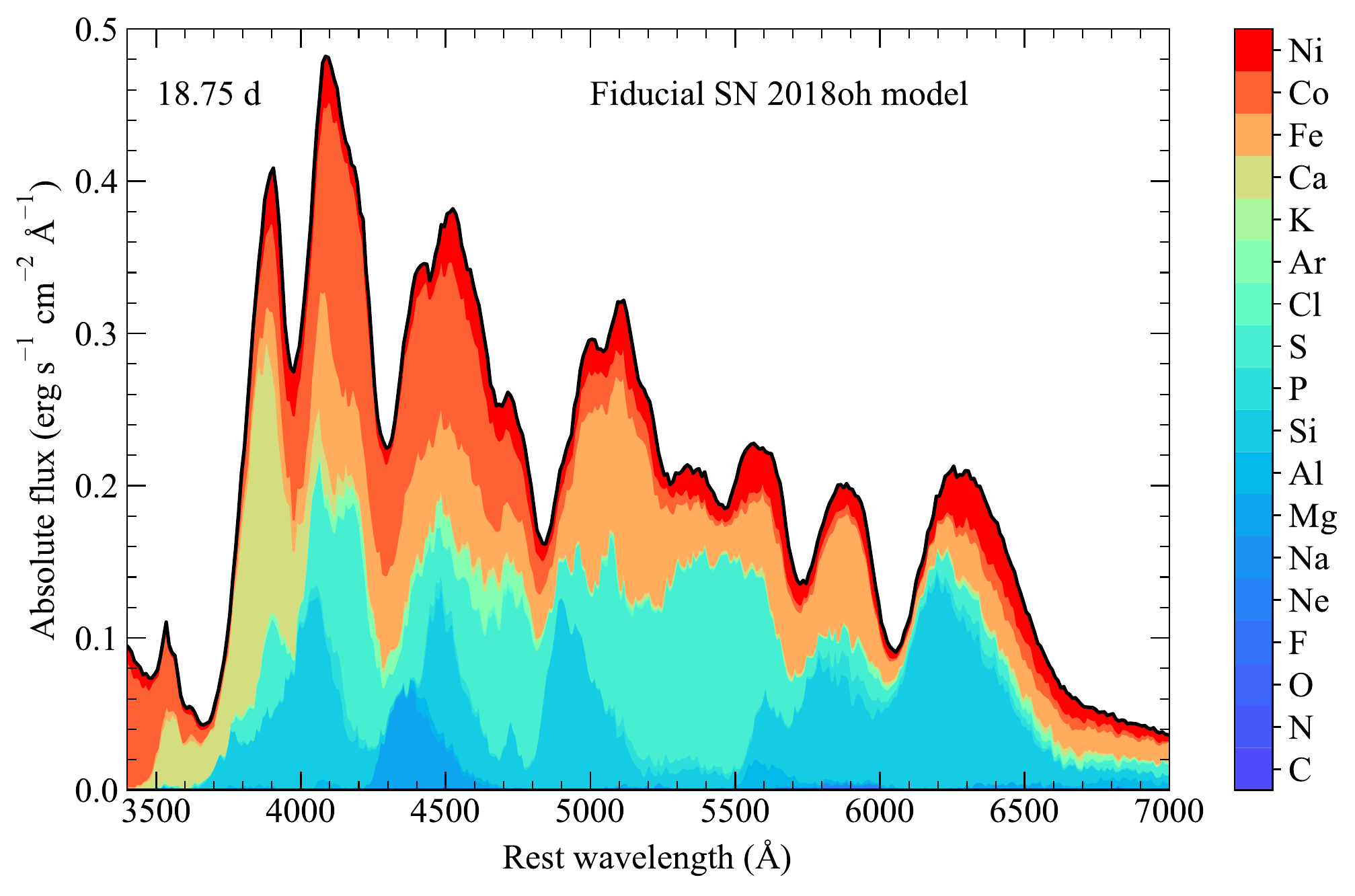}
    \end{subfigure}
    \begin{subfigure}[b]{0.49\textwidth}
        \includegraphics[height=5.9cm]{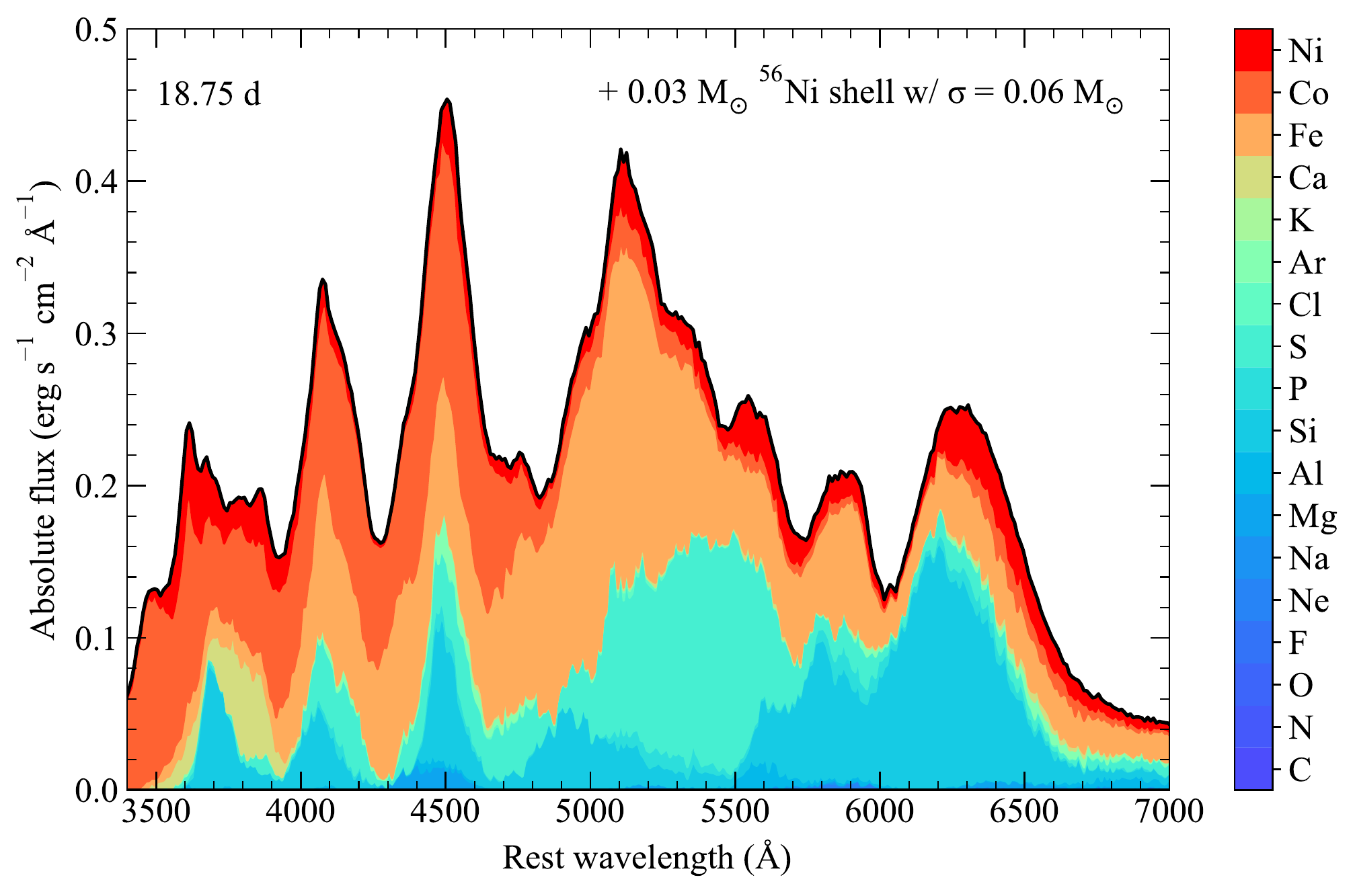}
    \end{subfigure}
    \caption{Contribution of individual elements within the model ejecta to the flux at each wavelength. Element fluxes are calculated based on their contribution to the opacity bin with which escaping Monte Carlo packets experienced their last interaction. Spectra are shown for our fiducial model ({\it left}) and our model that includes an additional $^{56}$Ni shell of 0.03~$M_{\rm{\odot}}$ and width of 0.06~$M_{\rm{\odot}}$ ({\it right}). Spectra are also shown for 1.50\,d after explosion, binned to $\Delta\lambda$ = 15~\AA\, ({\it top}), and 18.75\,d after explosion, binned to $\Delta\lambda$ = 10~\AA\, ({\it bottom}).
    }
    \label{fig:Kromer_plots}
\end{figure*}

In order to determine which elements are responsible for producing the features observed in the spectra, we tracked the opacity bin with which each packet experienced its last interaction and the contribution of each element to the total opacity within this bin. When packets were then binned in frequency to produce the spectrum, this gave the contribution of each element to the total luminosity at a given wavelength (see e.g. Fig.~2 of \citealt{artis} and Fig.~8 of \citealt{15h}). This is shown in Fig.~\ref{fig:Kromer_plots} for our 1.50\,d spectra, where each element is denoted by a different colour. Fig.~\ref{fig:Kromer_plots} shows that the feature around $\sim$4\,800~\AA\, could likely be attributed to \ion{Mg}{ii}. At all other wavelengths there does not appear to be a dominant source of opacity, hence the last element with which a packet experienced an interaction is approximately random and there is a lack of features in the spectrum. For our model with a 0.03~$M_{\rm{\odot}}$ $^{56}$Ni shell, Fig.~\ref{fig:Kromer_plots} shows that Ni is the dominant source of opacity at all wavelengths. This is unsurprising given that most packets escaping at this time will have originated from the $^{56}$Ni shell.

\par

Closer to maximum light, at 18.75\,d after explosion, differences in spectra between the models are more subtle, as demonstrated by Fig.~\ref{fig:clump_spectra}. We again note that limitations in the composition of our models can result in velocities of IMEs that are too high. For the models shown in Fig.~\ref{fig:clump_spectra}, the \ion{Si}{ii}~$\lambda$6\,355 velocity is $\sim$15\,000~km~s$^{-1}$ at maximum light, while the velocity in SN~2018oh is $\sim$10\,300~km~s$^{-1}$ \citep{li--19}. In Fig.~\ref{fig:Kromer_plots}, we also show the contribution of each element to the luminosity at every wavelength for our fiducial model and model with a 0.03~$M_{\rm{\odot}}$ $^{56}$Ni shell. For all models, we find that the \ion{Si}{ii}~$\lambda$6\,355 feature is unaffected by the presence of $^{56}$Ni shells. Indeed, wavelengths longer than $\sim$4\,500~\AA\, show only small variations at this time. A prominent exception to this is the \ion{S}{ii} `W' feature around $\sim$5\,400~\AA. Figure~\ref{fig:Kromer_plots} indicates that this feature has become affected by the presence of additional Fe in all but our fiducial model. The most noticeable differences for our model spectra around maximum light occur at wavelengths $\lesssim$4\,500~\AA. Figure~\ref{fig:Kromer_plots} shows that the strong \ion{Ca}{ii}~H\&K features are either significantly weakened or completely removed by the presence of additional IGEs. 

\par

Even for relatively small $^{56}$Ni shell of 0.01~$M_{\rm{\odot}}$ (representing $\lesssim$2\% of the total $^{56}$Ni mass for the models in Fig.~\ref{fig:18oh_clump_lc}), a dramatic increase in luminosity at early times can be achieved. Such signatures should be easily detected in observations of SNe~Ia, provided they are discovered sufficiently early (within $\sim$3 days of explosion).

\subsection{$^{56}$Ni shells versus extended $^{56}$Ni distributions}
\label{sect:clump_v_extended}
 
 In Fig.~\ref{fig:18oh_clump_lc} we also show comparisons of our $^{56}$Ni shell models to a model with a more extended $^{56}$Ni distribution relative to our fiducial model. In this case, the $^{56}$Ni distribution in the outer $\sim$0.1~$M_{\rm{\odot}}$ is comparable to our 0.01~$M_{\rm{\odot}}$ shell model with $\sigma$ = 0.18~$M_{\rm{\odot}}$. Deeper inside the ejecta however, the $^{56}$Ni distributions differ significantly. Figure~\ref{fig:18oh_clump_lc} clearly shows the consequences of these differences in $^{56}$Ni distributions. 
 
 \par 
 
 Our extended $^{56}$Ni distribution model shows a broad light curve that smoothly increases in brightness towards maximum light. For our 0.01~$M_{\rm{\odot}}$ shell model with $\sigma$ = 0.18~$M_{\rm{\odot}}$, the light curve is similar to that of the extended $^{56}$Ni distribution model until $\sim$1.5 days after explosion. After this point, the two models clearly diverge. The shell model shows a somewhat flattening of the rise between $\sim$1.5 -- 3 days, before subsequently rising more sharply again until maximum light. The colour evolution also differs significantly for both models, as shown in Fig.~\ref{fig:18oh_clump_lc}(d). The shell model is easily distinguished based on the sudden shift towards redder colours following explosion, while our extended $^{56}$Ni distribution model shows a relatively flat colour evolution during the first 10 days post-explosion.
 
 \par

The spectra of our extended $^{56}$Ni distribution model are also shown in Fig.~\ref{fig:clump_spectra}. At both early times and close to maximum light, the extended $^{56}$Ni distribution model appears similar to our 0.01~$M_{\rm{\odot}}$ $^{56}$Ni shell model. This further highlights the need for continuous follow up as both models are reasonably indistinguishable without observations towards the end of the bump phase ($\sim$2 -- 4\,d after explosion).

  \par
 
 As discussed in \cite{magee--20}, a light curve bump cannot be produced solely by having a large fraction of $^{56}$Ni in the outer ejecta. Our models clearly demonstrate that a bump in the light curve requires the $^{56}$Ni distribution to vary non-monotonically. A smoothly decreasing $^{56}$Ni fraction towards the outer ejecta will only vary the width and colours of the light curve, but does not produce a flux excess resembling those of models with $^{56}$Ni shells. If the shell is extended over a relatively large region of the ejecta ($\sim$0.18~$M_{\rm{\odot}}$), a broader light curve can be produced, but even in this case the model is clearly distinguished from a monotonically decreasing extended $^{56}$Ni distribution based on the colour evolution.

%

%

 %
\section{Comparisons with observations of SNe Ia}
\label{sect:comparisons}

\subsection{SN~2018oh}
\label{sect:comparisons_18oh}

\begin{figure*}
\centering
\includegraphics[width=\textwidth]{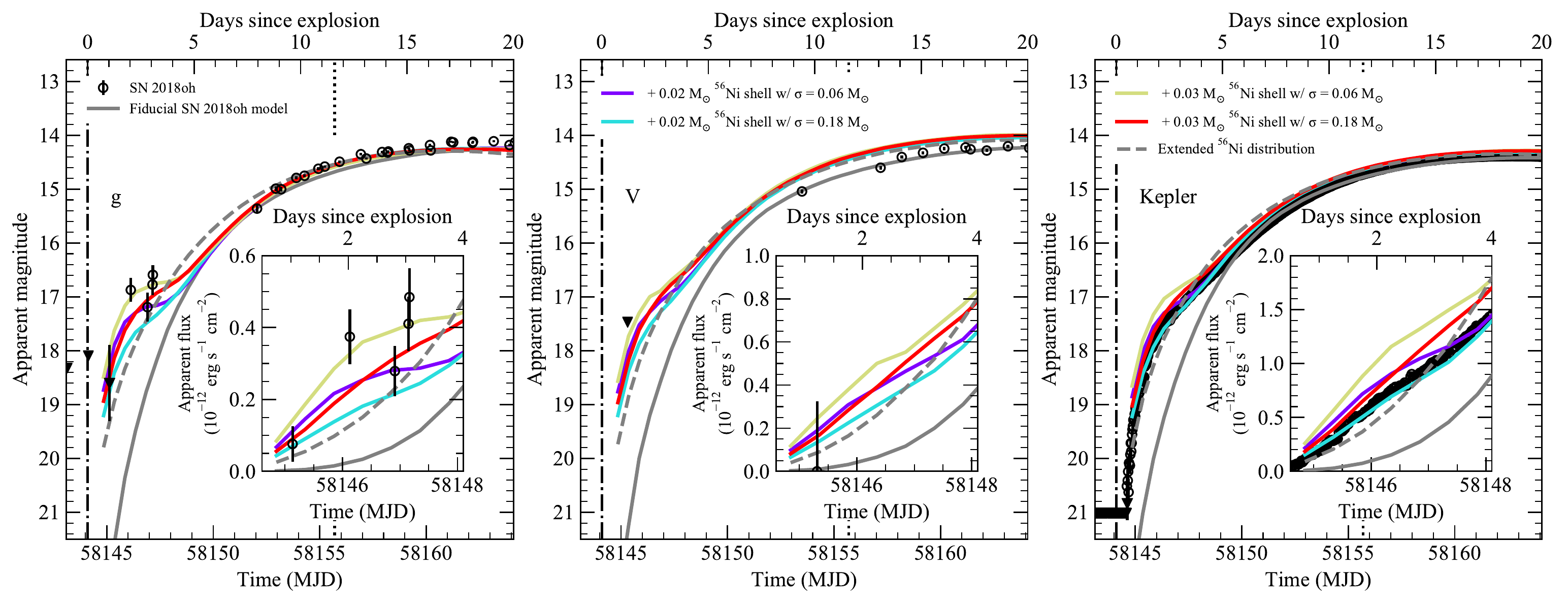}
\caption{Comparison of SN~2018oh to models with and without $^{56}$Ni shells in the outer ejecta. From left to right, panels shown the $g$-, $V$-, and Kepler-bands. Observations of SN~2018oh are shown as black points. Non-detections and detections of $\textless$3$\sigma$ are shown as downward triangles. Our fiducial model is shown as a grey solid line, while models with $^{56}$Ni shells are shown as coloured lines. Based on comparisons to our models, we infer an explosion date of MJD = 58144.1, which is shown as a vertical dash-dot line. A model with an extended $^{56}$Ni distribution relative to our fiducial SN~2018oh model is shown as a dashed line. The epoch of our spectral comparison is shown as a vertical dotted line. In each panel we show a zoom-in of the first four days after explosion. We note that this does not cover the epoch of the spectral comparison.}
\label{fig:18oh_lc_comp}
\centering
\end{figure*}

In the following section, we compare our model light curves and spectra to observations of SN~2018oh. We find the best agreement when assuming an explosion date of MJD = 58144.1 and distance modulus of $\mu$ = 33.66~mag. and adopt these values throughout this work. We note that the explosion date found here is 0.2\,d earlier than the first-light time derived by \cite{li--19} and the distance modulus is fully consistent. The spectrum of SN~2018oh has been corrected for galactic extinction only (host reddening is estimated to be negligible; \citealt{li--19}) and was obtained from WISeREP \citep{wiserep}. In order to ensure accurate flux calibration, we calculated synthetic magnitudes from the spectrum. We then calculated the offset relative to co-temporal photometric magnitudes and applied this across all bands using a third-order polynomial. The observed spectrum was then scaled to match the photometry in each band.

\par

In Fig.~\ref{fig:18oh_lc_comp}, we show a comparison between our models and observations of SN~2018oh. As demonstrated by Fig.~\ref{fig:18oh_lc_comp}, our fiducial model provides good agreement with the light curve beginning approximately one week after explosion, but is clearly significantly fainter than the observations at early times. Extending the $^{56}$Ni distribution of our fiducial model, such that $^{56}$Ni is present throughout the ejecta (see Fig.~\ref{fig:clump_distributions}), does not reproduce the shape of the light curve and instead results in a broader light curve that passes through the early flux excess. As in \cite{magee--20}, this indicates that the early excess is not simply due to a large fraction of $^{56}$Ni in the outer ejecta. 

\par

Rather than a simple extended $^{56}$Ni distribution, Fig.~\ref{fig:18oh_lc_comp} shows that models with $^{56}$Ni shells can broadly reproduce the light curve shape of SN~2018oh throughout its evolution to maximum light. The mass of the $^{56}$Ni shell required to match the flux excess of SN~2018oh is between 0.02 -- 0.03~$M_{\rm{\odot}}$. This is consistent with the value found by \cite{dimitriadis--19}, although we note that differences in the total $^{56}$Ni mass and distribution, and density profile compared to our models likely accounts for differences in the overall light curve shapes. Less massive $^{56}$Ni shells (0.01~$M_{\rm{\odot}}$) do not reach the required luminosities to match the flux excess around $\sim$2 days after explosion.    

\par

Aside from the mass of the $^{56}$Ni shell, there is some tension between the $g$- ($3\,800~\AA \lesssim \lambda \lesssim 5\,400~\AA$) and Kepler- ($4\,200~\AA \lesssim \lambda \lesssim 9\,000~\AA$) bands over the width of the shell that is required. We find that our model with a narrow ($\sigma = 0.06~M_{\rm{\odot}}$) 0.03~$M_{\rm{\odot}}$ shell produces the best match to the $g$-band light curve of SN~2018oh, while in the Kepler-band the model is somewhat too bright (by $\lesssim$0.5~mag.) during the excess phase. For the other models shown in Fig.~\ref{fig:18oh_lc_comp}, we find broad shells ($\sigma = 0.18~M_{\rm{\odot}}$) generally produce better agreement with the shape of the the Kepler light curve, but are slightly fainter than the $g$-band observations. In addition, our models with $^{56}$Ni shells are systematically brighter than the $V$-band observations around maximum light, while our fiducial model shows better agreement. 

\par

\begin{figure}
\centering
\includegraphics[width=\columnwidth]{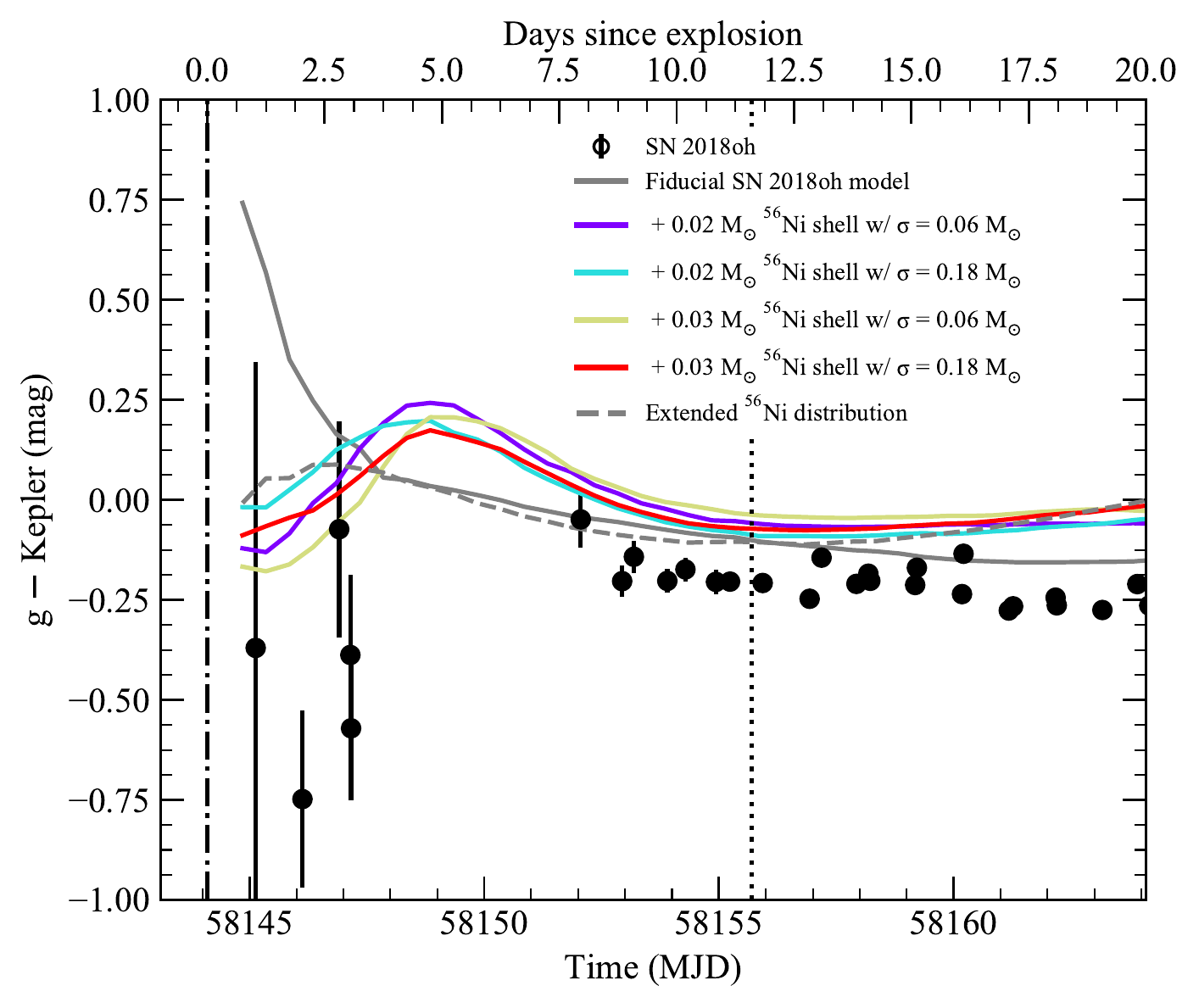}
\caption{Colour evolution of SN~2018oh compared to models with and without $^{56}$Ni shells in the outer ejecta. Observations of SN~2018oh are shown as black points. Our fiducial model is shown as a solid grey line. Models with additional $^{56}$Ni shells are shown as coloured lines and our model with an extended $^{56}$Ni distribution relative to our fiducial model is shown as a grey dashed line. Based on comparisons to our models, we infer an explosion date of MJD = 58144.1, which is shown as a vertical dash-dot line. The epoch of our spectral comparison is shown as a vertical dotted line.}
\label{fig:18oh_colour_comp}
\centering
\end{figure}

\begin{figure}
\centering
\includegraphics[width=\columnwidth]{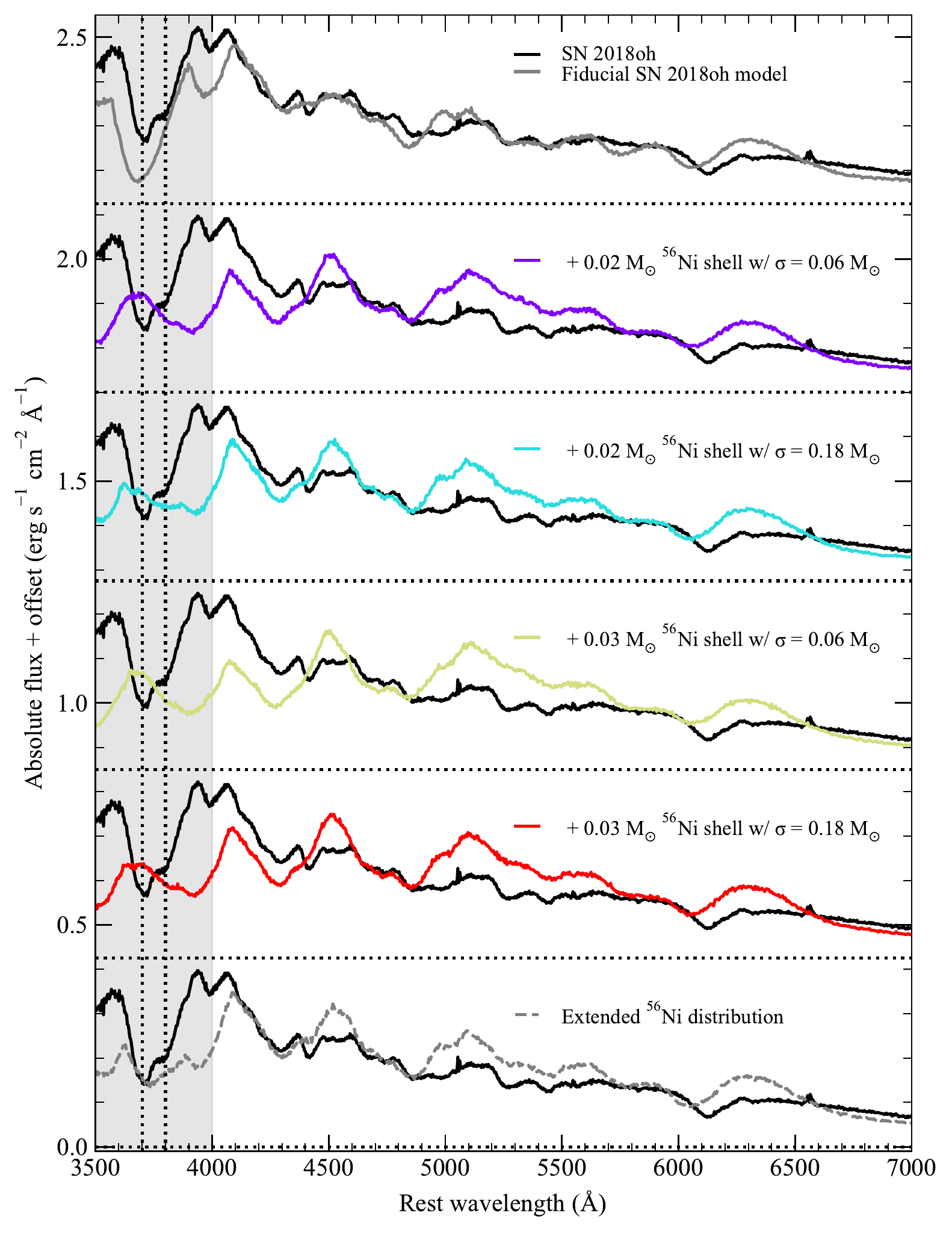}
\caption{Spectral comparison between SN~2018oh (black) and our models 11 days after explosion.  Each spectrum is offset vertically for clarity, with a dashed line showing the zero-point for each offset. The region showing strong flux suppression in our shell models is given by a shaded grey region. \ion{Ca}{ii}~H\&K and \ion{Si}{ii}~$\lambda$3858 lines are denoted by vertical dotted lines.   }
\label{fig:18oh_spec_comp}
\centering
\end{figure}

These discrepancies are further highlighted by the colour (Fig.~\ref{fig:18oh_colour_comp}) and spectral evolution (Fig.~\ref{fig:18oh_spec_comp}). Figure~\ref{fig:18oh_spec_comp} shows that even for our fiducial model there is disagreement between the model and observed spectra. This can be attributed to the simplified composition used which, as discussed in Sect.~\ref{sect:models} and Sect.~\ref{sect:clump_effects}, results in IME velocities that are too high. Regardless of these shortcomings, Fig.~\ref{fig:clump_spectra} shows that spectral features at longer wavelengths are largely unaffected by the presence of shells, while shorter wavelengths are much more sensitive. Therefore,
comparisons between our shell and fiducial models, and SN~2018oh, are still useful for inferring the effect of $^{56}$Ni shells on spectra, and investigating whether any specific signatures they might have at shorter wavelengths are present in SN~2018oh.

\par

Shortly after explosion, Fig.~\ref{fig:18oh_colour_comp} shows that our fiducial model is significantly redder than SN~2018oh, in addition to being fainter. The presence of the $^{56}$Ni shell provides an additional heating source in those models, and hence their bluer colours are more similar to SN~2018oh at this time (although we note the large uncertainties). Unfortunately, $g$-band observations are unavailable for the period MJD = 58148 -- 58151 ($\sim$3.8 -- 6.8\,d post-explosion), which coincides with the largest discrepancies in colour between our fiducial and $^{56}$Ni shell models.

\par

This difference in colour is further reflected in the spectrum at approximately 11.5\,d after explosion. For our fiducial model, although the region around $\sim$3500 -- 4000~$\AA$ (shaded in Fig.~\ref{fig:18oh_spec_comp}) is fainter than SN~2018oh, the overall shapes of the features are quite similar -- with the exception being that the \ion{Ca}{ii}~H\&K and \ion{Si}{ii}~$\lambda$3858 lines are blended in our model (dotted lines in Fig.~\ref{fig:18oh_spec_comp}). This is clearly not the case in all of our other models however, which show a significant amount of line blanketing. As discussed in Sect.~\ref{sect:model_lightcurves}, for all models with $^{56}$Ni shell there is a strong suppression of flux at wavelengths $\lesssim$4\,500~$\AA$, which can be attributed to the presence of additional IGEs. In addition, all models with $^{56}$Ni shells show a relatively flat feature between $\sim$3\,700 -- 4\,000~$\AA$ that is not seen in SN~2018oh, or indeed even the model with an extended $^{56}$Ni distribution.

\par

Following the colour evolution to maximum light, our shell models reach an inversion shortly before two weeks after explosion. At this point, the colour flattens before gradually becoming increasingly red. Again, this can be attributed to the presence of the $^{56}$Ni shell causing an increasing amount of blanketing as more flux emerges from the inner regions of the ejecta. Conversely, our fiducial model does not become as red as the shell models, although it remains redder than SN~2018oh by $\sim$0.05~mag.

\par

\par

Based on comparisons with observations of SN~2018oh, our models indicate that while $^{56}$Ni shells of $\sim$0.02 -- 0.03~$M_{\rm{\odot}}$ can reproduce the early light curve shape there is an overall negative effect on the spectra closer to maximum light. While some differences in features between our models and observations can be attributed to the simple composition used, a strong suppression of flux in the region $\sim$3\,700 -- 4\,000~$\AA$ appears to be a generic feature of models with relatively massive $^{56}$Ni shells (massive enough to match the early flux excess in the light curve). A similar feature is not observed in SN~2018oh and therefore would suggest that such shells are not present in the outer ejecta. Figure~\ref{fig:18oh_colour_comp} also clearly demonstrates the need for continuous and well-sampled multi-band follow-up of SNe~Ia, as the shape of the colour evolution can provide a key diagnostic between the different ejecta structures.

\subsection{SN~2017cbv}

\begin{figure*}
\centering
\includegraphics[width=\textwidth]{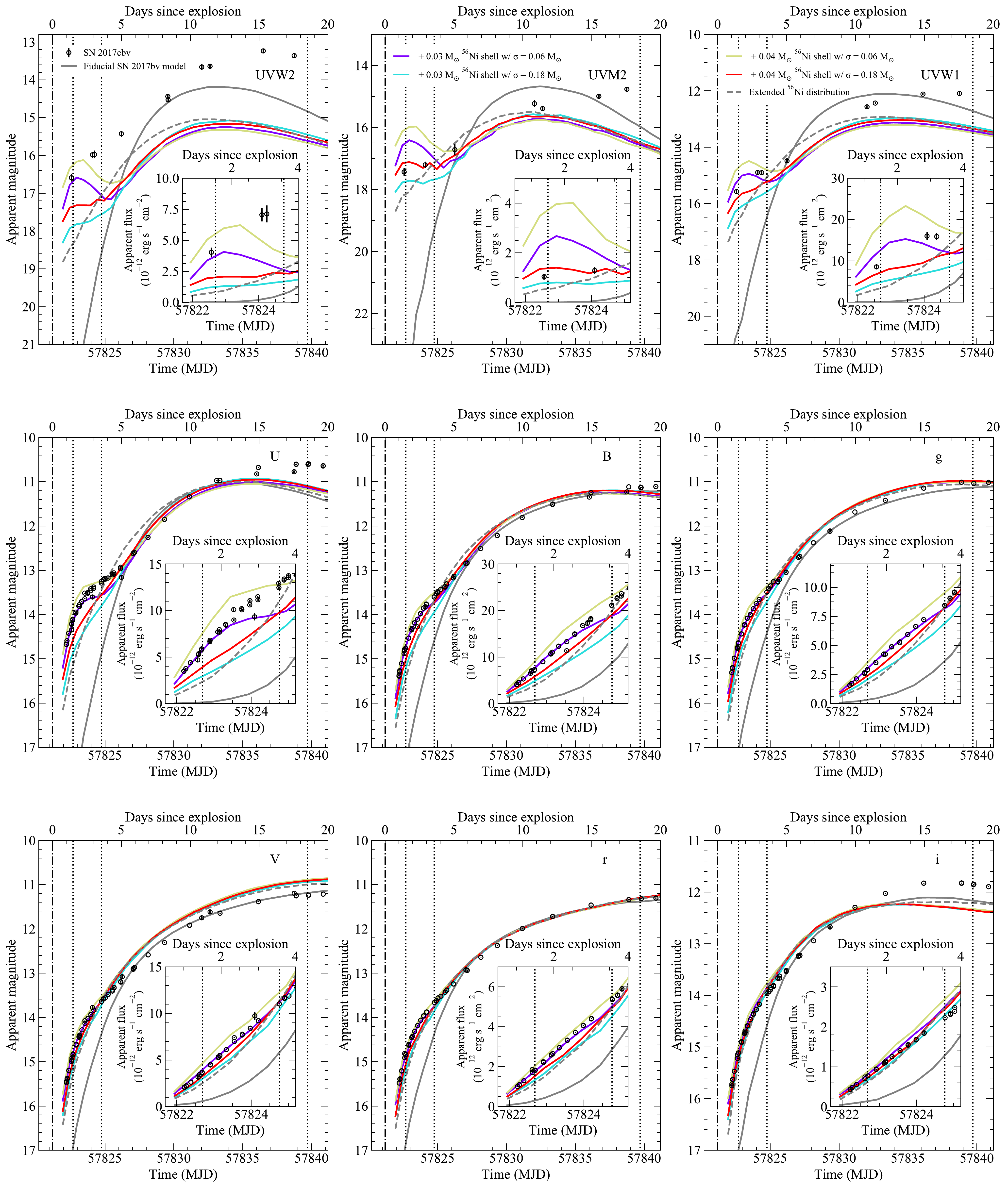}
\caption{Comparison of SN~2017cbv light curves to models with and without $^{56}$Ni shells in the outer ejecta. SN~2017cbv is shown as black points. The fiducial model for SN~2017cbv is shown as a solid grey line. Models with $^{56}$Ni shells added to the fiducial model are shown as coloured lines. Based on comparisons to our models, we infer an explosion date of MJD = 57821.2, which is shown as a vertical dash-dot line. A model with an extended $^{56}$Ni distribution relative to our fiducial SN~2017cbv model is shown as a dashed line. The epochs of our spectral comparisons are shown as vertical dotted lines. In each panel we show a zoom-in of the first four days after explosion.}
\label{fig:17cbv_lc_comp}
\centering
\end{figure*}

We now look to compare our models to observations of SN~2017cbv. Here we assume an explosion date of MJD = 57821.2 and distance modulus of $\mu$ = 30.74. The distance modulus used in this work is consistent with that of \cite{hosseinzadeh--17}, while the explosion date is 0.7 days earlier. Spectra of SN~2017cbv were corrected for galactic extinction only (host reddening is estimated to be negligible; \citealt{hosseinzadeh--17}), obtained from WISeREP \citep{wiserep}, and originally published in \cite{hosseinzadeh--17}. Spectra were flux calibrated in the same manner as described for SN~2018oh (see Sect.~\ref{sect:comparisons_18oh}).

\par

A comparison between the light curves of SN~2017cbv \citep{hosseinzadeh--17} and our models is shown in Fig.~\ref{fig:17cbv_lc_comp}. Figure~\ref{fig:17cbv_lc_comp} demonstrates that our fiducial model generally provides reasonable agreement with SN~2017cbv close to maximum light across the optical bands. For the UV bands ($UVW2$, $UVM2$, $UVW1$, and $U$) however, our fiducial model peaks earlier and fainter than the UV observations. As mentioned previously, we have assumed $^{56}$Ni constitutes 100\% of the total IGE immediately following explosion. The UV bands are more highly sensitive to the chosen metallicity of the ejecta. A more detailed composition and structure of the ejecta may result in improved agreement. The effect of metallicity on the model observables will be more fully explored in future work.

\par

Figure~\ref{fig:17cbv_lc_comp} shows that models with $^{56}$Ni shells in the outer ejecta can also provide reasonable agreement with the early light curve shape of SN~2017cbv. We find models with $^{56}$Ni shells of 0.03 -- 0.04~$M_{\rm{\odot}}$ provide the best match to the shape of the $BVgri$ band light curves within the days following explosion. The overall level of agreement in these bands is generally insensitive to the width of the $^{56}$Ni shell, but this is clearly not the case for the $U$-band. Models with narrow $^{56}$Ni shells ($\sigma$ = 0.06~$M_{\rm{\odot}}$) reproduce the shape of the $U$-band flux excess better than those with broader $^{56}$Ni shells ($\sigma$ = 0.18~$M_{\rm{\odot}}$), but even in these cases our models show an overall flatter flux excess than observed in SN~2017cbv. In addition, our models predict a strong UV bump within the first few days of explosion that is not observed in SN~2017cbv. Around maximum light, our $^{56}$Ni shell models show significant differences compared to our fiducial model and show a much closer resemblance to the extended $^{56}$Ni distribution model.

\par

\begin{figure*}
\centering
\includegraphics[width=0.8\textwidth]{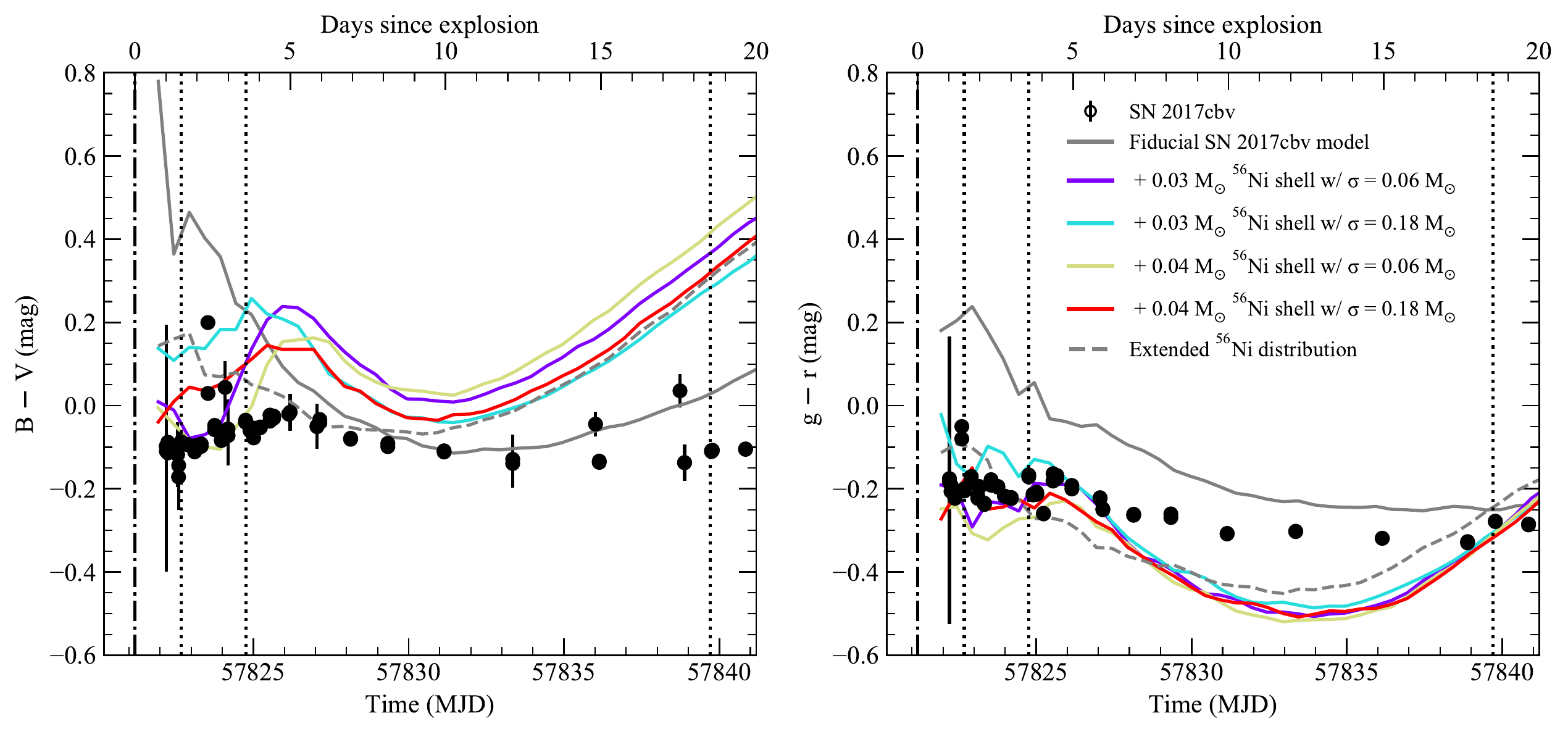}
\caption{Comparison of the $B-V$ (left) and $g-r$ (right) colour evolution in SN~2017cbv (black points) to models with (coloured lines) and without $^{56}$Ni shells (grey lines) in the outer ejecta. A model with an extended $^{56}$Ni distribution relative to our fiducial SN~2017cbv model is shown as a dashed line. The epochs of our spectral comparisons are shown as vertical dotted lines.}
\label{fig:17cbv_colour_comp}
\centering
\end{figure*}

\begin{figure*}
\centering
\includegraphics[width=\textwidth]{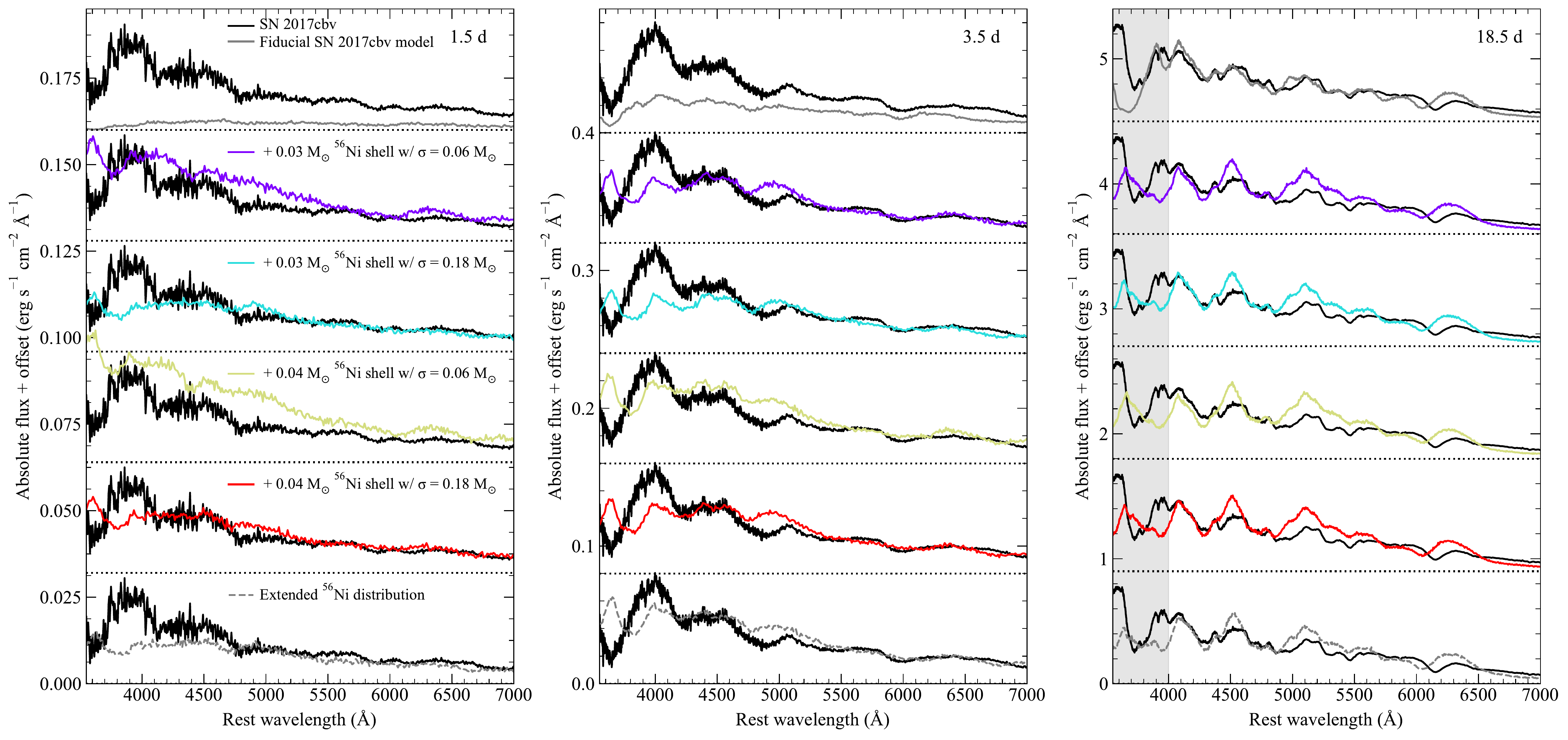}
\caption{Spectral comparison between SN~2017cbv (black) and our models. From left to right panels show spectra at 1.5, 3.5, and 18.5 days after explosion. Each spectrum is offset vertically for clarity, with a dashed line showing the zero-point for each offset. The region showing strong flux suppression at maximum light in our shell models is given by a shaded grey region. }
\label{fig:17cbv_spectra_comp}
\centering
\end{figure*}
   
Unlike SN~2018oh, SN~2017cbv has well sampled multi-band observations and spectra extending from explosion to after maximum light and therefore provides a better opportunity to determine which models can be excluded. Here, we consider spectra at three important epochs during the evolution of SN~2017cbv -- shortly after explosion (+1.5\,d), towards the end of the flux excess phase (+3.5\,d), and close to maximum light (+18.5\,d). As with SN~2018oh, we find that our fiducial model does not provide perfect agreement at maximum light. We again note that the effect of the shell is most pronounced at shorter wavelengths, while longer wavelengths are less affected (such as the \ion{Si}{ii}~$\lambda$6\,355 feature). We therefore focus our spectral comparisons on investigating signatures of the shells at shorter wavelengths.

\par

In Fig.~\ref{fig:17cbv_colour_comp} we show the colour evolution ($B-V$ and $g-r$) of SN~2017cbv compared to our models. Immediately following explosion, our fiducial model is clearly significantly redder than SN~2017cbv, and also shows no prominent features in its spectra (Fig.~\ref{fig:17cbv_spectra_comp}). Those models with narrow $^{56}$Ni shells ($\sigma$ = 0.06~$M_{\rm{\odot}}$) provide better agreement with the colour evolution during these epochs. Figure~\ref{fig:17cbv_spectra_comp} shows that at 1.5\,d after explosion the features around $\sim$3\,600 and 4\,200~\AA\, in SN~2017cbv are reasonably well matched by our narrow $^{56}$Ni shell models, although the models show a velocity that is too low. By 3.5\,d, we also find that the colours and spectroscopic features of some of our shell models are similar, although again the velocities of IGE features are too low in the model. Better agreement may be achieved if the density profile was altered such that the $^{56}$Ni shell was shifted to higher velocities. 

\par

After these early phases the initial light curve bump has subsided and the colour evolution of SN~2017cbv shows a small increase to redder colours until approximately one week after explosion. At this point the evolution turns towards bluer colours again. Our fiducial and extended $^{56}$Ni distribution models (i.e. those models without $^{56}$Ni shells) clearly do not match the shape of this colour evolution. Initially their colours are very red and become increasingly blue, and then steadily becoming redder again from approximately one week after explosion. In contrast, our $^{56}$Ni shell models show a similar colour inversion as in SN~2017cbv, although the colours are systematically shifted redder by $\sim$0.2 mag. Models with narrower $^{56}$Ni shells also typically show more dramatic changes in colour than SN~2017cbv. For example, over approximately five days, SN~2017cbv shows a shift of $\Delta (B-V) \sim 0.1$~mag. This is similar to the colour change in our broader $^{56}$Ni shell models, but significantly smaller than that of our narrow $^{56}$Ni shell models where $\Delta (B-V) \gtrsim 0.2$~mag. This is clearly in disagreement with the $U$-band light curve shape and the early spectral comparisons, for which we find better agreement with narrow $^{56}$Ni shell.

\par 

We note that the $g-r$ colour of SN~2017cbv does not show a similar colour evolution as in the $B-V$ colour during the early epochs. Instead the colour becomes slightly bluer (by $\sim$0.1~mag.) before flattening. In contrast, our $^{56}$Ni shell models still show a slight colour inversion. Additionally, the $g-r$ colours of these models are significantly bluer than SN~2017cbv beginning approximately one week after explosion. Instead, the shape of the colour evolution in our fiducial model is more similar to that of SN~2017cbv (although the colours are systematically redder in the model), excluding the initial few days during the flux excess phase.

\par

Although the early light curve shapes are reasonably well reproduced in our $^{56}$Ni shell models, it is clear that at maximum light there is significant disagreement, as shown in Fig.~\ref{fig:17cbv_spectra_comp}. As in SN~2018oh, we find that close to maximum light, our models do not reproduce the \ion{Ca}{ii}~H\&K and \ion{Si}{ii}~$\lambda$3\,858 lines. Instead, our models show strong flux suppression in this region due to the presence of additional IGEs. Although at early times it is significantly fainter than SN~2017cbv, our fiducial model provides the best agreement around maximum. As in the case of SN~2018oh, our models show that the flux suppression in the near-UV appears to be a generic feature of models with $^{56}$Ni shell. Those models with shells large enough to match the light curve shape all exhibit this feature, which is not observed in the spectra of SN~2017cbv. Therefore, we argue that while a $^{56}$Ni shell could explain the observed light curve shape, it is inconsistent with the spectral evolution. 



\section{Discussion}
\label{sect:discussion}

\par

Our models show that $^{56}$Ni shells in the outer ejecta can produce an excess of flux during the first few days after explosion -- the strength and duration of which depend on both the mass and width of the shell. Comparing our models to observations of SNe~2018oh and 2017cbv, we find that massive ($\sim$0.02 -- 0.04~$M_{\rm{\odot}}$) $^{56}$Ni shells can in general reproduce the shape of the optical light curve bumps and colours within the first few days of explosion. Such massive shells produce strong flux suppression at near-UV wavelengths however, which is not seen in the spectra of either object. In Sect.~\ref{sect:sources} we first discuss predictions of $^{56}$Ni shells from existing explosion models, while in Sect.~\ref{sect:alternatives} we discuss alternative scenarios for producing early light curve bumps and how they compare to those produced by $^{56}$Ni shells.

\subsection{Sources of non-monotonic $^{56}$Ni distributions in the outer ejecta}
\label{sect:sources}

As discussed in Sect.~\ref{sect:models}, our one-dimensional shell models may be considered qualitatively similar to multi-dimensional models in which an excess of $^{56}$Ni is produced directly along the line of sight. In other words, the one-dimensional shell is equivalent to a localised clump of $^{56}$Ni. Many of the models proposed for SNe~Ia invoke deflagrations to some degree, which could provide one method of producing $^{56}$Ni clumps \citep{khokhlov--93, hoeflich--96a, gamezo--05, bravo--06, jordan--08}. Generally, in such models the interior of the white dwarf is initially burned by a sub-sonic deflagration. As the burning front propagates, it will eventually become subject to Rayleigh-Taylor instabilities that wrinkle and deform the flame \citep{khokhlov--94, khokhlov--95}. At the same time, buoyancy of the hot, burned ash accelerates the deflagration plumes towards the surface of the white dwarf \citep{reinecke--02b}. The overall effect of the deflagration is irregular or uneven burning, and deflagration plumes could therefore potentially create $^{56}$Ni clumps in the outer ejecta. Multi-dimensional simulations in general however, do not show strong variations in the light curves as a function of viewing angle, which may be expected if there were significant differences in composition along certain directions (see e.g. \citealt{roepke--07, seitenzahl--13, sim--13, fink-2014, long--14}).

\par

One potential scenario for producing large $^{56}$Ni clumps in the outer ejecta lies in   gravitationally confined detonations \citep[GCDs; ][]{plewa--04}. This scenario does produce an excess that is qualitatively similar to SNe~2018oh and 2017cbv, although much less extreme. Here, the deflagration plumes reach the surface of the white dwarf before burning transitions to a detonation. As the plumes travel along the surface, they eventually converge on the stellar point opposite that of the initial breakout and it is this convergence that triggers the detonation. The result is a highly asymmetric ejecta composition in which clumps of deflagration ash are present at high velocities. \cite{seitenzahl--16} present light curves and spectra for one realisation of the GCD scenario, which shows a strong dependence on viewing angle. When viewed close to the direction of the deflagration ignition, a small bump ($\lesssim$0.5 mag. and lasting $\sim$3 days) in the $U$-band light curve is observed (see Fig.~7 of \citealt{seitenzahl--16}), resulting from the presence of the deflagration ash. Whether the GCD scenario could produce larger $^{56}$Ni clumps similar to those studied here remains to be seen, but warrants further investigation. 

\par

We note that it has also been argued that the GCD scenario may not be robustly achieved \citep{roepke--off-centre} and while initial models designed to mimic the GCD scenario could reproduce observables of SNe~Ia \citep{kasen--05}, subsequent multi-dimensional explosion simulations have shown good agreement with neither normal nor 91T-like SNe Ia \citep{seitenzahl--16}. In addition, the abundance distribution appears dissimilar to that inferred for normal SNe~Ia based on late-time spectra \citep{maguire--18}, as much of the stable material produced in the high density core would be carried to the outer ejecta through buoyancy.

\par

Aside from models invoking deflagrations, the double-detonation scenario is another possibility for producing large $^{56}$Ni mass fractions in the outer ejecta \citep{livne--90, woosley--94, bildsten--07}. In these models, a thin helium shell builds up on the surface of the WD. Depending on certain conditions (e.g. mass of the WD, mass of the helium shell, etc.), the shell may ignite and create a secondary detonation that leads to the complete disruption of the star \citep{fink--07}. In general, burning in the helium shell can produce a large amount of $^{56}$Ni, qualitatively similar to the $^{56}$Ni shells included in this work \citep{fink--10, shen--10, moll--13}. Similar to our models, the presence of a $^{56}$Ni shell produces an excess of flux within the days following explosion. The duration and luminosity of this flux excess is affected by both the mass and post-explosion composition of the helium shell, and is comparable to that predicted by our $^{56}$Ni shell models \citep{maeda--18,polin--19}. Although some double-detonation models predominantly produce $^{56}$Ni during the shell burning, other short-lived radioactive isotopes are also synthesised -- in particular $^{48}$Cr and $^{52}$Fe \citep{woosley--94}. Aside from affecting the light curves, the presence of these additional isotopes also produces a distinctive \ion{Ti}{ii} ($^{48}$Ti is a daughter product of the $^{48}$Cr decay chain) trough between $\sim$4\,000 -- 4\,300~$\AA$. Even for models in which the helium shell products are dominated by $^{56}$Ni, a strong \ion{Ti}{ii} feature is produced \citep{polin--19}. The presence of such a feature could therefore provide one method for distinguishing a $^{56}$Ni excess produced by helium shell detonations and other methods. Finally, \cite{maeda--18} present spectra calculated from models with and without helium burning ash, which are qualitatively similar to our models with $^{56}$Ni shells and fiducial models. Unlike our models, \cite{maeda--18} find that spectra of at least some models with helium shells containing $^{56}$Ni are redder at three days after explosion than those models without shells. This behaviour is opposite to that of our $^{56}$Ni shell models and likely results from the increased line blanketing due to additional elements within the outer ejecta, rather than solely $^{56}$Ni as in our models.

\par

Even relatively small $^{56}$Ni shells in the outer ejecta (0.01~$M_{\rm{\odot}}$ of $^{56}$Ni above a mass coordinate of 1.35~$M_{\rm{\odot}}$) can have a significant effect on the light curve, producing a light curve bump that increases the bolometric magnitude by $\textgreater$2 magnitudes at one day after explosion. Given that such bumps are not observed for the majority of SNe~Ia, this would indicate that $^{56}$Ni shells (at least those large enough to reproduce the light curve shapes of SN~2018oh and SN~2017cbv) must be relatively rare phenomena, if indeed they occur at all.

\subsection{Alternative sources of flux excesses}
\label{sect:alternatives}

While our models show that $^{56}$Ni shells in the outer ejecta can produce an excess of flux at early times, similar behaviour is found for other scenarios that do not invoke the presence of additional $^{56}$Ni -- in particular, interaction between the SN ejecta and companion star or circumstellar material (CSM). In both cases, this material is shock heated by the SN ejecta and gradually cools, providing an additional source of luminosity.

\par

\cite{kasen--10} shows how the total luminosity produced during the light curve bump varies depending on the nature of the companion, with larger mass and more evolved companions producing larger flux excesses.  For red giant companions, the $B$-band flux excess is approximately one magnitude fainter than peak, while in the ultraviolet it is approximately one magnitude brighter. This is unlike our $^{56}$Ni shell models, in which the flux excess is always $\gtrsim$2 mag. fainter than peak. In contrast, less evolved main sequence stars show flux excesses covering a similar range in magnitude to that of our $^{56}$Ni shell models, although they are generally shorter lived. For example, when assuming interaction occurs with a 6~$M_{\rm{\odot}}$ MS companion, the $B$-band light curve has already plateaued within $\sim$1 day of explosion. The companion interaction scenario has been proposed for the light curve bumps in SNe~2018oh and 2017cbv, and is the favoured interpretation in both cases \citep{hosseinzadeh--17, dimitriadis--19, shappee--2019}. We note, however, that both SNe show a fainter UV luminosity than predicted by this model and show no signs of hydrogen, which is expected to be swept up during the interaction and produce strong emission at late times \citep{sand--18, tucker--19}. 

\par

An early light curve bump may also be produced due to the interaction between the SN ejecta and CSM. This CSM may result from either accretion of a non-degenerate companion or following a merger. In the case of SN~2017cbv however, the presence of CSM from a non-degenerate companion was investigated by \cite{ferretti--17}, who find no evidence for significant amounts of material. Light curve bumps resulting from post-merger CSM interactions were investigated by \cite{piro-16}, who show that various light curve bumps can be produced depending on the mass and radial extent of the material. \cite{piro-16} also demonstrate that the distribution of $^{56}$Ni within the ejecta plays an important role in shaping the light curve bump, as those models with more extended $^{56}$Ni distributions can mask the signatures of interaction. In all cases, those CSM models without mixed $^{56}$Ni distributions produce pronounced bumps in the $V$-band light curve that differ significantly from the shapes of the bumps in our models. For those models with more extended $^{56}$Ni distributions, the CSM mass must be relatively large ($\sim$0.3~$M_{\rm{\odot}}$) to produce a flux excess as long lived as our $^{56}$Ni shell models.

\par

Within the merger scenario, an accretion disk may also be formed around the more massive primary \citep{levanon--15}. Based on analytical models, \cite{levanon--17} show that following the explosion this disk-originated matter (DOM) is shock heated and produces early UV emission. \cite{levanon--19} argue that the early light curve bump of SN~2018oh is consistent with interaction between SN ejecta and DOM. The timescale and luminosity of the flux excess produced by the DOM interaction scenario is also similar to some our of $^{56}$Ni shell models, and therefore warrants further investigation with full radiative transfer calculations.

%

\section{Conclusions}
\label{sect:conclusions}

Following from the works of \cite{dimitriadis--19} and \cite{shappee--2019}, we investigated $^{56}$Ni shells in the outer ejecta as a source of the flux excess observed in SNe~2018oh and 2017cbv. For each object, we took a fiducial model that reproduced the later light curve (i.e ignoring the early flux excess) and added $^{56}$Ni shells to investigate the changes relative to a model without a shell. We presented light curves and spectra for models containing $^{56}$Ni shells of 0.01 -- 0.04~$M_{\rm{\odot}}$. In all cases, we found that models with $^{56}$Ni shells produce an excess of flux during the first few days after explosion. Even a relatively small $^{56}$Ni shell of 0.01~$M_{\rm{\odot}}$ can have a profound effect on the early light curve and produce an increase in the bolometric magnitude \textgreater2 mag. at one day after explosion. The appearance of the flux excess is also affected by the shape of the $^{56}$Ni shell, with narrower shells producing excesses with a more well-defined peak than broader shells. In addition, our models show that the effects of the $^{56}$Ni shells are most prominent at shorter wavelengths and the colour evolution of models with $^{56}$Ni shells differs significantly. All models with $^{56}$Ni shells show an inversion in the $B-V$ colour evolution between $\sim$3 -- 6 days after explosion and are clearly distinguished from the fiducial models and even models with more extended $^{56}$Ni distributions.

\par

We compared the light curves of our models with $^{56}$Ni shells to observations of SNe~2018oh and 2017cbv and found improved agreement compared to the fiducial models -- our $^{56}$Ni shell models can produce the overall light curve shape in both objects. Similar to \cite{dimitriadis--19}, we found that a $^{56}$Ni shell of 0.02 -- 0.03~$M_{\rm{\odot}}$ is required to reproduce the shape of SN~2018oh, while for SN~2017cbv a slightly higher mass of 0.03 -- 0.04~$M_{\rm{\odot}}$ is required. Although our $^{56}$Ni shell models can reproduce the early optical light curve bumps of both objects, there are noticeable differences in the colours and spectral evolution. Our models are systematically redder and show more extreme changes in colour than the observations. A strong suppression of flux in the region of $\sim$3\,700 -- 4\,000~$\AA$ also appears to be a generic feature in the maximum light spectra of our $^{56}$Ni shell models. Neither SN~2018oh nor SN~2017cbv show a similar feature, indicating that while models with $^{56}$Ni shells could match the light curve shapes, the spectra are in disagreement. In addition, our $^{56}$Ni shell models produce a strong bump in the UV light curves that is not observed. Overall, this would suggest that $^{56}$Ni shells are not the source of the early excess flux in either object.  

\par

Based on existing explosion models, producing such shells (or clumps) in the outer ejecta as required also appears challenging. The GCD scenario can produce $^{56}$Ni clumps in the outer ejecta for certain viewing angles and hence can also produce a small bump in the light curve. Such clumps however, are not as extreme as would be required to match the light curve shapes of SNe~2018oh and 2017cbv. Whether this scenario could produce larger $^{56}$Ni clumps and good agreement with observations remains to be seen, but warrants further investigation. Alternatively, the double detonation scenario can also produce large $^{56}$Ni fractions in the burned ash of the accreted helium shell. Again, this scenario can also produce light curve bumps, but the picture is complicated by the presence of short-lived isotopes providing an additional powering source for the early light curve. In addition, the helium shell ash produces distinct spectroscopic features, such as strong \ion{Ti}{ii} absorption. The presence of such a feature could therefore provide one method of distinguishing between $^{56}$Ni shells or clumps produced by either deflagration or helium shell ash.

\par

Studies of SNe Ia have shown that early light curve bumps are rare. Only a handful of objects displaying bumps are currently known, despite extensive searches. Based on a sample of 35 SNe Ia with sufficiently early light curves, \cite{magee--20} find $\sim$20 -- 30\% of objects show evidence for a flux excess when compared against their model light curves. Given that even a relatively small $^{56}$Ni shell can significantly alter the light curve shape, this would indicate that non-monotonic $^{56}$Ni distributions in the outer ejecta (at least those similar to what is required to match the light curves of SNe~2018oh or 2017cbv) must also be relatively rare -- if they occur naturally at all. Future surveys dedicated to the discovery and follow-up of SNe at early times will help to shed light on the nature of flux excesses in SNe~Ia. 

%

\begin{acknowledgements}
We thank M. Bulla and S. A. Sim for their help in implementing virtual packets into TURTLS. We thank the referees for their suggestions to improve the clarity of the paper. This work was supported by TCHPC (Research IT, Trinity College Dublin). Calculations were performed on the Kelvin cluster maintained by the Trinity Centre for High Performance Computing. This cluster was funded through grants from the Higher Education Authority, through its PRTLI program. This work made use of the Queen's University Belfast HPC Kelvin cluster. MM and KM are funded by the EU H2020 ERC grant no. 758638. This research made use of \textsc{Tardis}, a community-developed software package for spectral synthesis in supernovae
\citep{tardis}. The development of \textsc{Tardis} received support from the
Google Summer of Code initiative and from ESA's Summer of Code in Space program. \textsc{Tardis} makes extensive use of Astropy and PyNE.
\end{acknowledgements}

\bibliographystyle{aa}
\bibliography{Magee}

\end{document}